\documentstyle[12pt,epsf]{article}
\begin{document}
\title{Resummation of anisotropic quartic oscillator.
Crossover from anisotropic to isotropic large-order behavior}
\author{
\footnotesize H.\ Kleinert and S.\ Thoms \\
\footnotesize Institut f\"ur Theoretische Physik \\
\footnotesize Freie Universit\"at Berlin \\
\footnotesize Arnimallee 14 \\
\footnotesize 14195 Berlin
\and
\footnotesize W.\ Janke \\
\footnotesize Institut f\"ur Physik, Johannes \\
\footnotesize Gutenberg-Universit\"at Mainz \\
\footnotesize Staudinger Weg 7 \\
\footnotesize 55099 Mainz}
\date{ }
\maketitle
\begin{abstract}
We present an approximative calculation of the ground-state energy for
the anisotropic oscillator with a potential
\begin{displaymath}
V(x,y)=\frac{1}{2} (x^2+y^2)+\frac{g}{4}  \left[x^4+
2(1-\delta)x^2 y^2+y^4 \right]. 
\end{displaymath} 
Using an instanton solution of the isotropic action $\delta = 0$, 
we obtain the imaginary part of the ground-state energy for small
negative $g$ as
a series expansion in the anisotropy parameter $\delta$. From
this, the large-order behavior of the $g$-expansions accompanying each
power of $\delta$ are obtained by means of a dispersion relation in $g$.
These $g$-expansions are summed by a Borel transformation,
yielding an approximation to the ground-state energy for the region near the
isotropic limit. This approximation is found to be
excellent in a rather wide region of $\delta$ around $\delta = 0$.

Special attention is devoted to the immediate vicinity of the isotropic point.
Using a simple model integral we show that the large-order behavior
of an $\delta$-dependent series expansion in $g$ undergoes a crossover 
from an isotropic to an anisotropic regime as the order $k$ of the expansion
coefficients passes the value $k_{{\rm cross}} \sim 1/ |{\delta}|$.    
\end{abstract}
\section{Introduction} 
Phase transitions in anisotropic systems with cubic symmetry have 
attracted much interest in the literature \cite{Aharon}-\cite{KleSchu2}. 
Especially well studied are corresponding models in quantum mechanics.
To gain an analytic insight into the 
latter, Banks, Bender, and Wu (BBW) \cite{BaBeWu} studied a Hamiltonian with 
a potential 
\begin{equation}
\label{pot}
V = \frac{g}{4} \left[x^4+2(1-\delta)x^2 y^2+y^4 \right].
\end{equation} 
Using multidimensional WKB techniques they derived the
large-order behavior of the perturbation series for the ground-state
energy
\begin{equation}
\label{perturb}
E=\sum_k E_k(\delta) \, g^k  
\end{equation}
as a function of the anisotropy parameter $\delta$.

In $1990$, Janke \cite{Jan} derived the same results with more efficiency 
from a path integral for the imaginary part of the energy $E$.
The imaginary part contains information on the tunneling decay rate of the
ground state for $g < 0$, and 
determines directly the large-order behavior
of the perturbation coefficients via a dispersion relation in the complex
coupling constant plane.
Both, BBW and Janke, find a different
large-order behavior of the isotropic system $\delta = 0$ and the anisotropic 
system $\delta \neq 0$. They do not discuss, however, the interesting
question of how the latter goes over into the former as $\delta$ goes to zero.
 
It is the purpose of this paper to fill this gap.
For an optimal understanding of the expected behavior we shall not attack
directly the path integral involving the potential (\ref{pot}),
but first only the corresponding simple integral. 
For this integral, a perturbation expansion of the form (\ref{perturb})
yields exactly-determined $\delta$-dependent perturbation coefficients.
The coeffients $E_k(\delta)$ are shown to have a large-order behavior 
which undergoes a crossover between the earlier derived isotropic and 
anisotropic behaviors when the order $k$ passes the crossover value
$k_{{\rm cross}} \sim 1/ |{\delta}|$.

The expansion terms of a model integral with the potential
(\ref{pot}) counts the number of terms in a perturbation expansion of the  
quantum mechanic and the field theory. Thus the bare model integral is
sufficient to derive nontrivial information on the large-order behavior of the
eventual object of interest, quantum field theory. It turns out that for 
resumming the $g$-series, asymptotic large-order estimates for the 
$\delta$-dependent coefficients can be used only in the anisotropic 
regime $k |{\delta}| \gg 1$. In the isotropic regime $k |{\delta}| \ll 1$, on
the other hand, it is impossible to truncate the large-order expansion
of the perturbation coefficients after a finite number of terms. Thus the
neighbourhood of the isotropic system $\delta = 0$ needs an extra 
investigation. In the context of quantum field theory, this was recently 
delivered in \cite{KleTho}. 

The imaginary parts of physical quantities at small negative $g$  
can be calculated with the help of classical solutions called instantons. 
In systems sufficiently close to the isotropic point it is not necessary to 
know the exact instanton solutions for all $\delta$. The knowledge of the 
solution at the symmetry point $\delta = 0$ is perfectly sufficient, around 
which the imaginary parts can be expanded in powers of $\delta$.

After having understood the model integral, we shall perform the same 
analysis for an anisotropic quantum mechanical system, which represents an
one-dimensional $\phi^4$-field theory with cubic anisotropy.

The paper is organized as follows. In \mbox{Section $2$} we develop a
simple resummation procedure by which the divergent power series expansion
of a function $Z(g)=\sum_k Z_k \, g^k$ is converted into an almost
convergent series $\sum_p a_p \, I_p (g)$. Here $I_p (g)$ are certain 
confluent hypergeometric functions which possess power
series expansions in $g$ with similar large-order behavior as the system 
under study. 
In \mbox{Section $3$} we shall analyse the above-mentioned crossover in 
the large-order behavior for the simple model integral. 
In particular, we shall justify the resummation procedure 
of \mbox{Section $2$} and the methods in \cite{KleTho} to be the perfect
tools in approximating the integral for the region near the isotropic 
limit $\delta \rightarrow 0$. In \mbox{Section $4$}, finally, we present
a similar calculation for the ground state energy of the anharmonic
potential with cubic anisotropy. 

In addition to this more standard resummation procedure we analyse the model
also within the variational perturbation theory developed 
in \cite{FeKle}-\cite{KaKle}. 
Variational perturbation theory yields uniformily and exponentially fast 
converging expansion for quantum mechanical systems with quartic 
potentials \cite{JaKle}. 
The uniform convergence was first proven for the partition function of the 
anharmonic integral, later for the quantum mechanical anharmonic 
oscillator with coupling strength $g$ in several papers \cite{SerConv}.
Recently, the proof was sharpened and extended to the energies \cite{CoKleJa}.

The input for the quantum mechanical model is provided by the exact 
Rayleigh-Schr{\"o}dinger perturbation coefficients of the ground-state energy
which we derive from an extension of recursion relations first shown by 
Bender and Wu (BW) \cite{BeWu}.  
\section{Resummation}
We begin with developing a practical algorithm for a Borel resummation
of a divergent perturbation series
\begin{equation}
\label{res1}
Z(g)=\sum_k Z_k \, g^k \, .
\end{equation}
Our method will be most efficient under the following conditions:   
\begin{enumerate}
\item From low-order perturbation theory we know the expansion
      coefficients $Z_k$ up to a certain finite order $N$.
\item From semiclassical methods we are in the possession of the
      high-order information in the form
      \begin{equation}
      \label{res2}
      Z_k\stackrel{k \rightarrow \infty}{\longrightarrow}\,
      \gamma \left(-1\right)^k k!\, k^{\beta} {\sigma}^k 
      \left(1+\frac{{\gamma}_1}{k}+\frac{{\gamma}_2}{k}+ \cdots \right). 
      \end{equation}
\item By some scaling arguments we are able to assure a power behavior
      in the strong coupling limit
      \begin{equation}
      \label{res3}
      Z(g)\stackrel{g \rightarrow \infty}{\longrightarrow}\,
      \kappa g^{\alpha}
      \end{equation}
\end{enumerate}
The idea of the algorithm is the following: \\
It must be possible to construct an infinite, complete set of Borel
summable functions $I_p (g)$ which satisfy the high-order and strong-coupling
conditions (\ref{res2}) and (\ref{res3}). These functions
can be used as a new basis in which to reexpand $Z(g)$:
\begin{equation} 
\label{res4}
Z(g)=\sum_{p=0}^{\infty} a_p \, I_p (g) \, ,
\end{equation}
The series (\ref{res4}) should be such that the knowledge of the first
$(N+1)$ coefficients in the power series expansion (\ref{res1}) is
sufficient to determine directly the first $(N+1)$ coefficients $a_p$, 
yielding an approximation
\begin{equation} 
\label{res5}
Z(g) \approx Z^{(N)}(g) \equiv \sum_{p=0}^N a_p \, I_p (g)\, .
\end{equation}
This would then be a new representation of the function $Z(g)$ 
with the same power series up to $g^N$ but which makes use of
large-order and strong-coupling informations (\ref{res2}) and (\ref{res3}). 
In the limit of large $N$,
the series (\ref{res5}) is expected to converge towards the exact solution.

The functions $I_p (g)$ being Bore summable have a Borel representation
\begin{equation} 
\label{res6}
I_p (g)=\int_0^{\infty} dt e^{-t} t^{b_0} B^{b_0}_p (gt)\, .
\end{equation}
Parametrized by some $b_0$ and integer $p$, what are the conditions 
on $B_p(gt)$, such that $I_p (g)$ satisfies (\ref{res2}) and (\ref{res3}) 
for all $p$?
The answer is most easily found with the help of the hypergeometric functions
\begin{equation}
\label{res7}
_2F_1(a,b;c;-\sigma gt)=\sum_{k=0}^{\infty}\frac{(a)_k (b)_k}{(c)_k}
                  \frac{\left(-\sigma gt\right)^k}{k!}
\end{equation}
with appropriate parameters $a(p)$, $b(p)$ apd $c(p)$.
The Pochhammer symbol $(a)_k$ stands short for
$(a)_k=\Gamma(a+k)/{\Gamma(a)}$. These functions have the following
virtues: First, they are standard special functions of
mathematical physics whose properties are well-known. Second, they have 
a cut running from $t=-1/ |\sigma g|$ to minus
infinity which is necessary to generate the large-order behavior (\ref{res2}).
Third, they have enough free parameters to fit all input-data.
The first property permits an immediate calculation of the Borel 
integral (\ref{res6}), which is simply a Laplace transformation of
$t^{b_0}\, _2F_1(a,b;c;-\sigma gt)$
\begin{equation}
\label{res8}
\int_0^{\infty} dt e^{-t} t^{b_0}\, _2F_1(a,b;c;-\sigma gt)=
\frac{\Gamma(c)}{\Gamma(a)\Gamma(b)} E(a,b,b_0+1:c:1/{\sigma} g)
\end{equation}
The resulting $E(a,b,b_0+1:c:1/{\sigma} g)$ is MacRobert's $E$-function.
Using its asymptotic expansion (see Ref. \cite{Harm2}, page 203) 
it is easy to verify that our ansatz reproduces the large-order 
behavior (\ref{res2}).
Indeed, for large $k$ the power series
\begin{equation}
\label{res9}
\frac{\Gamma(c)}{\Gamma(a)
\Gamma(b)} E(a,b,b_0+1:c:1/{\sigma} g) \equiv \sum_{k=0}^{\infty} e_k \, g^k
\end{equation}
has coefficients which grow like
\begin{equation}
\label{res10}
e_k\stackrel{k \rightarrow \infty}{\longrightarrow}\,
\frac{\Gamma(c)}{\Gamma(a)\Gamma(b)}\left(-1\right)^k
k!\, k^{a+b-c+b_0-1} {\sigma}^k \, .
\end{equation}
Moreover, this property is unchanged if the original
hypergeometric function is multiplied by a power $(\sigma gt)^p$.  
A possible set of Borel functions are therefore the following functions: 
\begin{equation}
\label{res11}
B_p(gt)=(\sigma gt)^p \, _2F_1(a,b;c;-\sigma gt)\, .
\end{equation}
Looking at (\ref{res10}) we see that the functions (\ref{res11}) are not
completely fixed by a given large-order behavior. The parameter $\beta$
in (\ref{res2}) merely imposes the following relation upon the
parameters $a$,$b$,$c$ and $b_0$
\begin{equation}
\label{res12}
a+b-c+b_0-1=\beta \, ,
\end{equation}
and there are many different ways to satisfy this.
The specific choice will be suggested by practical considerations. 
One such consideration is that
the $I_p$'s should possess a simple integral representation in order to 
avoid complicated numerical work.
In addition, we would like to work with parameters $a$,$b$ and $c$, for which
the hypergeometric function $_2F_1$ reduces to simple
algebraic functions. This happens only for special sets of the parameters.
A simple possibility is for instance (see Ref. \cite{Abram}, page 556)
\begin{equation}
\label{res13} 
_2F_1(a,a+\frac{1}{2};2a+1;-z)=4^a \left(1+\sqrt{1+z}\right)^{-2a}
\end{equation}
which arises by choosing the parameters $a$,$b$ and $c$ which are related by
\begin{equation}
\label{res14}
a+b-c=-\frac{1}{2} \, ,  \quad ; \quad c-2b=0 \, .
\end{equation}
With this, the relation (\ref{res12}) can be satisfied for an
arbitrary value of the parameter $a$ by choosing
\begin{equation}
\label{res15}
b_0=\beta+\frac{3}{2} \, ,
\end{equation}
and we are left with only one parameter degree of freedom. This freedom
may be used to accommodate the strong-coupling behavior of $Z(g)$ if it 
is known. The equation (\ref{res5}) yields the condition 
$I_p(g)\rightarrow {\rm const}. \times g^{\alpha}$ on the functions $I_p(g)$.
From (\ref{res6}) we see that such a power behavior emerges if all 
Borel functions $B_p$ satisfy $B_p(z) 
\rightarrow {\rm const}. \times z^{\alpha}$ 
and thus $_2F_1(a,b;c;-z)\rightarrow {\rm const}. \times z^{-p+\alpha}$ 
[see (\ref{res11})]. 
The explicit representation (\ref{res13}) shows that the parameter $a$ has to 
be taken as
\begin{equation}
\label{res16}
a=p-\alpha \, .
\end{equation}
Thus we obtain the approximation $Z^{(N)}=\sum_{p=0}^N a_p \, I_p$ with
\begin{eqnarray}
\label{res17}
I_p(g)\!\!\!\! &=&\!\!\!\!\! \int_0^{\infty}dt 
\frac{e^{-t} t^{b_0}}{\Gamma(b_0+1)}
\frac{(\sigma gt)^p}{4^p}\, _2F_1\left(p-\alpha,p-\alpha+
\frac{1}{2};2(p-\alpha)+1;-\sigma gt\right) \nonumber \\
\!\!\!\! &=& \!\!\!\!\! \int_0^{\infty}dt 
\frac{e^{-t} t^{b_0}}{\Gamma(b_0+1)}
\left(\frac{1}{2}+\frac{1}{2}\sqrt{1+\sigma gt}\right)^{2\alpha}
\frac{(\sigma gt)^p}{\left(1+\sqrt{1+\sigma gt}\right)^{2p}} \,\, ,
\end{eqnarray} 
where the Borel parameter $b_0$ is fixed by (\ref{res15}). 
The normalization constant $1/4^p \Gamma(b_0+1)$ in front of the expansion
functions was introduced for convenience. 

Let us now derive equations for the expansion coefficients $a_p$ in terms
of the perturbation coefficients $Z_k$.
All one has to do is take the asymptotic expansions 
\begin{equation}
\label{res18}
I_p(g)=\sum_{k=0}^{\infty} I_k^p \, g^k \, ,
\end{equation}
insert these into (\ref{res5}), collect terms of equal power $g^k$, and 
compare these with the perturbation series (\ref{res1}). This gives the
$(N+1)$ algebraic equations
\begin{equation}
\label{res19}
Z_k^{(N)}\equiv\sum_{p=0}^N I_k^p \, a_p=Z_k\quad ; \quad k=0,1,\ldots, N
\end{equation}
By assumption, the series on the left hand side contains only the
coefficients $a_p$ with $p \leq N$. Thus the $a_p$'s can be computed,
in principle, by inverting the $(N+1) \times (N+1)$ matrix
$(I)_{kp}=I_k^p$. Even though this can be done recursively for any given 
case, it is preferable to find an explicit algebraic solution for $a_p$
in terms of $Z_k$. This is possible using the following trick.
We rewrite the asymptotic expansion of $Z^{(N)}$ in Borel form
\begin{equation}
\label{res20}
Z^{(N)}(g)\equiv \sum_{p=0}^N a_p \, I_p(g)=
\sum_{k=0}^{\infty} Z_k^{(N)} \, g^k = \int_0^{\infty} dt e^{-t} t^{b_0}
\sum_{k=0}^{\infty}\frac{Z_k^{(N)} (gt)^k}{\Gamma(k+b_0+1)} \, ,
\end{equation}
insert the expression (\ref{res17}) for $I_p(g)$, and compare directly
both integrands
\begin{equation}
\label{res21}
\sum_{p=0}^{N}a_p \frac{\left(\frac{1}{2}+\frac{1}{2}
\sqrt{1+\sigma gt}\right)^{2\alpha}}{\Gamma(b_0+1)}\left[
\frac{\sigma gt}{\left(1+\sqrt{1+\sigma gt}\right)^2}\right]^p=
\sum_{k=0}^{\infty} \frac{Z_k^{(N)}(gt)^k}{\Gamma(k+b_0+1)} \, .
\end{equation}
Introducing the new variable
\begin{equation}
\label{res22}
w\equiv\frac{\sigma gt}{\left(1+\sqrt{1+\sigma gt}\right)^2}=
\frac{\sqrt{1+\sigma gt}-1}{\sqrt{1+\sigma gt}+1} \, ,
\end{equation}
we obtain from (\ref{res21}) the relation valid for all $\alpha$:
\begin{equation}
\label{res23}
\sum_{p=0}^{N} a_p \, w^p=\sum_{k=0}^{\infty}\frac{Z_k^{(N)}}{
(b_0+1)_k}\left(\frac{4}{\sigma}\right)^k 
\frac{w^k}{(1-w)^{2(k-\alpha)}} \, .        
\end{equation}
In order to compare equal powers in $w$ we expand on the right hand side
\begin{equation}
(1-w)^{-2(k-\alpha)}=\sum_{l=0}^{\infty} 
\left(\begin{array}{c}
        -2(k-\alpha) \\ l
       \end{array} \right) (-w)^l 
\end{equation} 
which gives after a shift of the summation 
index from $l$ to $p=k+l$ the first $(N+1)$
coefficients $a_p$ in terms of the perturbation coefficients $Z_k$ 
\begin{equation}
\label{res24}
a_p=\sum_{k=0}^p \frac{Z_k}{(b_0+1)_k} \left(\frac{4}{\sigma}\right)^k
\left(\begin{array}{c}
        -2(k-\alpha) \\ p-k
       \end{array} \right) (-1)^{p-k} 
\end{equation}
(recall that $Z_k^{(N)}=Z_k$ for $k=0,1,\ldots, N$).
Finally, rewriting the binomial coefficients by means of the identity
\begin{equation}
\left(\begin{array}{c}
        x \\ p
       \end{array} \right)=(-1)^p
\left(\begin{array}{c}
        p-x-1 \\ p
       \end{array} \right) \, ,  
\end{equation}
we obtain the more convenient expression
\begin{equation}
\label{res25}
a_p=\sum_{k=0}^p \frac{Z_k}{(b_0+1)_k} \left(\frac{4}{\sigma}\right)^k
\left(\begin{array}{c}
        p+k-1-2\alpha \\ p-k
       \end{array} \right) \, .
\end{equation}
Thus, we have solved the original matrix inversion problem
(\ref{res19}) by translating it to a simple problem in function theory,
namely that of inverting the function $w(\sigma gt)$ in (\ref{res22}).
For the purpose of calculating the integrals $I_p(g)$ numerically,
we may use the variable $w$ itself as a variable
of integration, and rewrite the integral representation for $I_p(g)$ 
in the form: 
\begin{equation}
\label{res26}
I_p(g)=\left(\frac{4}{\sigma g}\right)^{b_0+1} \int_0^1 dw
\frac{
(1+w)w^{b_0+p}}{\Gamma(b_0+1) (1-w)^{2b_0+2\alpha+3}} 
\exp\left[-\frac{4w}{(1-w)^2 \sigma g}\right] \, .
\end{equation}
Together with the explicit formula (\ref{res25}) for the 
coefficients $a_p$ we thus have solved the resummation, and it is
now straightforward to calculate the approximation (\ref{res5}).
\section{Model integral}
In order to set up an approximation method for an anisotropic model in the
neighborhood of the isotropic point $\delta =0$, it is instructive to 
study first a simple toy model whose partition function is defined by a 
two-dimensional integral:
\begin{equation}
\label{igral1}
Z=\frac{1}{2\pi}\int\!\!\!\int_{-\infty}^{+\infty} dxdy 
\exp\left\{-\frac{1}{2}\left(x^2+y^2\right)-\frac{g}{4}\left[x^4+
2(1-\delta)x^2 y^2 +y^4\right]\right\} . 
\end{equation} 
This can be interpreted as a partition function of a $\phi^4$-theory
in zero spacetime dimensions with cubic anisotropy. Introducing 
polar coordinates
$x=r\cos\varphi$ and $y=r\sin\varphi$, we obtain the more convenient form of
the integral (\ref{igral1}):
\begin{equation}
\label{igral2}
Z=\frac{1}{2\pi}\int_0^{\infty}\int_0^{2\pi} d\rho d\varphi
\exp\left[-\rho-G(g,\delta,\varphi){\rho}^2 \right]
\end{equation}
with $\rho=r^2/2$ and
\begin{equation}
\label{igral3}
G(g,\delta,\varphi)=g\left[1-\frac{\delta}{2} \sin^2(2\varphi)\right] \, .
\end{equation}
After an integration over the angle $\varphi$, we find the integral
\begin{equation}
\label{igral4}
Z=\int_0^{\infty} d\rho \exp\left[-\rho-g\left(1-\frac{\delta}{4}\right)
{\rho}^2 \right] I_0\left(\frac{\delta}{4} g {\rho}^2\right) \, ,
\end{equation}
where $I_0(x)$ is a modified Besselfunction $I_{\nu}(x)$ for $\nu=0$.
Eq.(\ref{igral4}) is useful for a numerical calculation of 
$Z(g,\delta)$. It will serve as a testing ground for our approximations.

Thanks to the special spacetime dimensionality of the model, the perturbation 
expansion of $Z(g,\delta)$ can be obtained explicitely and we can calculate the
large-order behavior without doing the saddle point approximation, which is
unavoidable in quantum mechanics and field theory.

At first glance it seems useful to expand $Z(g,\delta)$ in the form
\begin{equation}
\label{igral5}
Z(g,\delta)=\sum_{k=0}^{\infty} Z_k(\delta) g^k \, ,
\end{equation}
where the perturbation coefficients are parametrized by the anisotropy
$\delta$.
The coefficients $Z_k(\delta)$ may be found by expanding the integrand
of (\ref{igral2}) in powers of $g$ and performing the integral term by
term:
\begin{equation}
\label{igral6}
Z_k(\delta)=\frac{(-1)^k}{k!}
\Gamma(2k+1)\left(1-\frac{\delta}{2}\right)^{k/2}P_k\left(\frac{4-\delta}{
2\sqrt{4-2\delta}}\right)\, ,
\end{equation}
where $P_k(x)$ are the Legendrepolynomials.

\begin{figure}[p]
\unitlength1cm
\vspace{-8ex}
\begin{minipage}[t]{13.2cm}
\epsfxsize=13cm
\hfill
\epsfbox{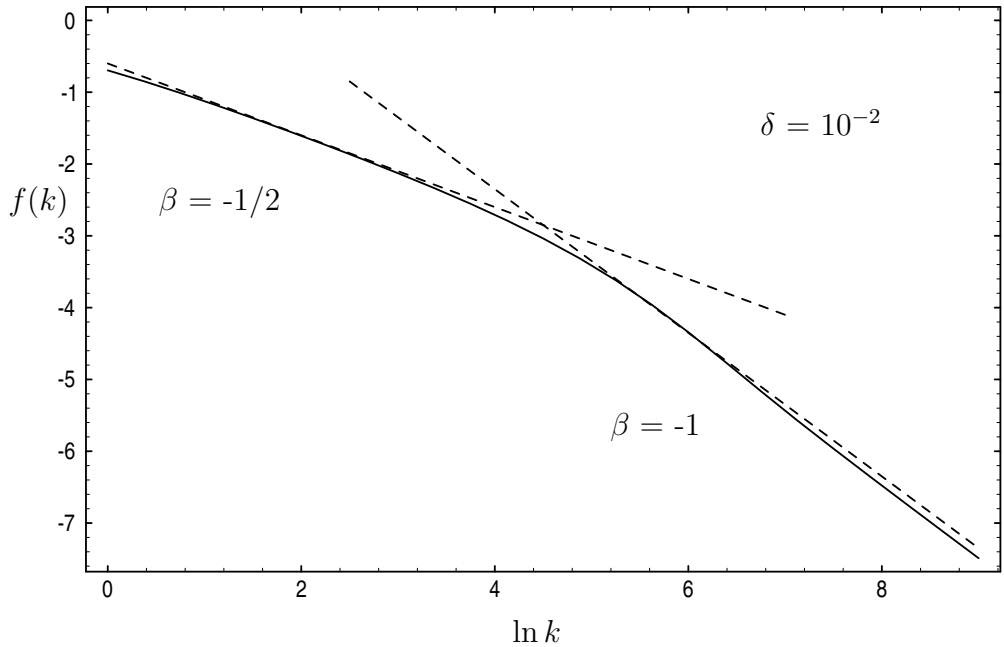} \par
\begin{picture}(13,2)
\put(0,10.0){$f(k)$}
\put(2.0,10.0){$\beta$ = -1/2}
\put(8.0,7.0){$\beta$ = -1}
\put(10.0,11.0){$\delta$ = $10^{-2}$}
\put(6.7,4.25){$\ln k$}
\end{picture}
\end{minipage}
\vspace{-20ex}  
\caption{\label{cr_ov2} Crossover of large-order behavior of the
expansion cofficients $Z_k$ in Eq. (\ref{igral5}) from the isotropic
regime $(\beta=-1/2)$ to the anisotropic regime $(\beta=-1)$. Plotted
is the function $f(k)=\ln \left[Z_k/(-4)^k k! \right]$ for the anisotropy
$\delta=10^{-2}$. In this case the crossover value is given
by $k_{{\rm cross}} \sim 1/ |{\delta}|=10^2$ 
$(\ln k_{{\rm cross}} \approx 4.6)$.}
\end{figure}

\begin{figure}[p]
\unitlength1cm
\vspace{-8ex}
\begin{minipage}[t]{13.2cm}
\epsfxsize=13cm
\hfill
\epsfbox{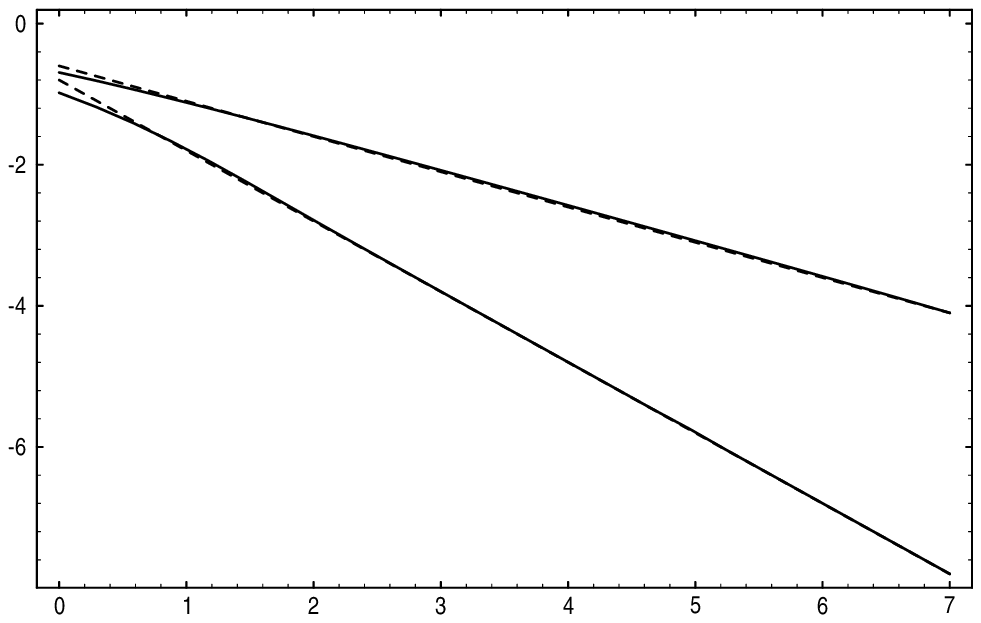} \par
\begin{picture}(13,2)
\put(0,9.5){$f(k)$}
\put(6,11.5){$a)$\ $\delta$ = $10^{-4}$ , $\quad \beta$ = -1/2} 
\put(1.7,8.5){$b)$\  $\delta$ = 1.0 , $\quad \beta$ = -1}
\put(6.7,4.25){$\ln k$}
\end{picture}
\end{minipage}
\vspace{-20ex}  
\caption{\label{cr_ov1} Example for the two different
large-order regimes of $Z_k$ in Eq. (\ref{igral5}), where $f(k)$ is the
same function as in Fig. \ref{cr_ov2}. $a)$: Isotropic regime 
($k_{{\rm {\rm cross}}}=10^4$, $\beta=-1/2$). $b)$: Anisotropic regime 
($k_{{\rm {\rm cross}}}=1$, $\beta=-1$).    }
\end{figure}

In Figure \ref{cr_ov2} we have plotted the order dependence of these
coefficients for the anisotropy parameter $\delta = 10^{-2}$. What we can
see is a crossover of the large-order behavior from an isotropic to an
anisotropic regime in the vicinity of an special crossover value 
$k_{{\rm cross}} \sim 1/ |{\delta}|=10^2$. In the anisotropic regime
$k |{\delta}| \gg 1$, the large-order parameter $\beta$ has the value
$\beta=-1$. For $k |{\delta}| \ll 1$, however, we can read off the large-order
behavior of the isotropic case, i.\ e.\ $\beta = -1/2$ (see also 
Figure \ref{cr_ov1}).

Using the duplication formula for Gamma-functions and the expansion
\begin{equation}
\label{igral7}
\Gamma(k+\varepsilon+1)=k!k^{\varepsilon}[1+{\cal O} (1/k)] \, ,
\end{equation}
the large-$k$ behavior of $\Gamma(2k+1)$ is given by
\begin{equation}
\label{igral8}
\Gamma(2k+1)=\pi^{-1/2} 4^k (k!)^2 k^{-1/2}[1+{\cal O} (1/k)]\, .
\end{equation}
An approximation of the Legendre Polynomials $P_k(x)$ for large $k$ including
contributions of the order ${\cal O} (1/k)$ can be derived from 
Hobson (see Ref. \cite{Hobson}, page 305):
\begin{eqnarray}
\label{igral9}
P_k(x)\!\!\!&=&\!\!\!(2\pi k)^{-1/2}\left(x^2-1\right)^{-1/4}
\left(x+\sqrt{x^2-1}\right)^{k+1/2} \nonumber \\
            & &\times \left[1+\frac{1-\sqrt{x^2-1}\left(x+
\sqrt{x^2-1}\right)}{8k\sqrt{x^2-1}\left(x+\sqrt{x^2-1}\right)}+
{\cal O}(1/k^2)\right] \, .
\end{eqnarray}
Substituting
\begin{equation}
x=\frac{4-\delta}{2\sqrt{4-2\delta}} \, ,
\end{equation}
we obtain for $\delta > 0$
\begin{eqnarray}
\label{igral10}
\lefteqn{P_k\left(\frac{4-\delta}{2\sqrt{4-2\delta}}\right)=} \nonumber \\
 & & \sqrt{\frac{2}{\pi}}k^{-1/2}
     {\delta}^{-1/2}\left(\frac{1}{1-{\delta}/2}\right)^{k/2}
     \left[1+\frac{1}{k\delta}\left(\frac{4-3\delta}{8}\right)+{\cal O}
     \left(\frac{1}{k^2 {\delta}^2}\right) \right] \, .
\end{eqnarray}
The combination of (\ref{igral8}) and (\ref{igral10}) yields the large-order
behavior of the perturbation coefficients $Z_k(\delta)$:
\begin{equation}
\label{igral11}
Z_k(\delta)=\frac{2^{1/2}}{\pi}(-1)^k 4^k k! k^{-1} {\delta}^{-1/2}
\left\{1+\frac{1}{k\delta}\left[\frac{1}{2}+{\cal O}(\delta)\right]+
{\cal O} \left( \frac{1}{k^2 {\delta}^2} \right) \right\} \, .
\end{equation}
A similar calculation can be done for $\delta < 0$ with the result:
\begin{eqnarray}
\label{igral1b2}
Z_k(\delta)\!\!\!&=&\!\!\!
\frac{(2-\delta)^{1/2}}{\pi}(-1)^k (4-2\delta)^k k! k^{-1} 
({-\delta})^{-1/2} \nonumber \\
& & \times
\left\{1-\frac{1}{k\delta}\left[\frac{1}{2}+{\cal O}(\delta)\right]+
{\cal O} \left( \frac{1}{k^2 {\delta}^2} \right) \right\} \, .
\end{eqnarray}
For resumming the series (\ref{igral5}), the perturbation 
coefficients (\ref{igral11}) and (\ref{igral1b2}) can be used only for 
$k|{\delta}| \gg 1$. In the regime $k|{\delta}| \ll 1$,
on the other hand, it is unpossible to truncate the series in (\ref{igral11})
and in (\ref{igral1b2}) after a finite order of $1/k|{\delta}| \gg 1$.
Thus, the isotropic regime cannot be described by
resumming the perturbation series (\ref{igral5}) using the asymptotic
results (\ref{igral11}) and (\ref{igral1b2}). 
Being interested in the region close to 
the isotropic limit, we therefore use an expansion different 
from (\ref{igral5}), and rewrite $Z(g,\delta)$ as
\begin{equation}
\label{igral12}
Z(g,\delta)=\sum_{k=0}^{\infty}\sum_{n=0}^k Z_{kn} \, g^k {\delta}^n \, .
\end{equation} 
Then, for the reason given in \cite{KleTho}, 
reasonable results should be obtained
by resumming the $g$-series accompanying each power ${\delta}^n$. The explicit
form of the coefficients $Z_{kn}$ is given by 
\begin{equation}
\label{igral13}
Z_{kn}={\pi}^{-1/2}(-1)^{k+n}\frac{\Gamma\left(n+\frac{1}{2}\right)
\Gamma(2k+1)}{2^n \Gamma(n+1)^2 \Gamma(k-n+1)} \, ,
\end{equation}
and $Z_{kn}=0$ for $k<n$.
For these coefficients, the expansion of the Gamma-functions yields the 
behavior for large $k \gg n$: 
\begin{equation}
\label{igral14}
Z_{kn}=(-1)^n \frac{\Gamma\left(n+\frac{1}{2}\right)}{
2^n \Gamma(n+1)^2} (-1)^k \frac{4^k}{\pi} k! k^{n-1/2}\left[
1+{\cal O}(1/k) \right] \, . 
\end{equation}

In the following we shall calculate the coefficients (\ref{igral14}) by means
of the steepest descent method using the saddle points of the limit 
$\delta \rightarrow 0$. This will serve as a simple preparation for
the analogous method in quantum mechanics and field theory.

\begin{figure}[p]
\unitlength1cm
\begin{center}
\begin{picture}(13,7)
\put(1.0,1.5){\epsfxsize=4cm \epsfbox{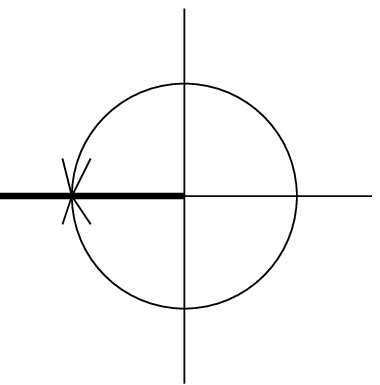}}
\put(7.0,1.5){\epsfxsize=5cm \epsfbox{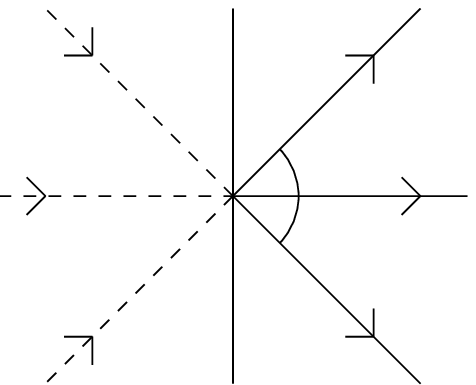}} 
\put(4.5,5.2){\circle{0.6}}
\put(4.4,5.1){$g$}
\put(2.8,0.5){$a)$}
\put(1,3.8){$+\pi$}
\put(1,3.1){$-\pi$}
\put(10,5.2){\circle{0.6}}
\put(9.9,5.1){$r$}
\put(9.3,0.5){$b)$}
\put(10.2,3.9){$+{\pi}/4$}
\put(10.2,2.9){$-{\pi}/4$}
\put(11.8,5.5){${\Gamma}_{-}$}
\put(11.8,3.7){${\Gamma}_{0}$}
\put(11.8,1.5){${\Gamma}_{+}$}
\end{picture}
\end{center}
\caption{\label{grplane} Analytic continuation
$g\rightarrow |g| \exp(\pm \pi)$: 
$a)$ Rotation by angles $\pm \pi$ in the cut complex $g$-plane. 
$b)$ Two rotated paths of integration in the $r$-plane $(r>0)$. }
\end{figure}

As a function of a $\delta$ and a complex coupling $g$,
the integral (\ref{igral1}) is defined in the 
half-plane ${\rm Re} \, g \geq 0$. 
For ${\rm Re} \, g <0 $, the integral can be calculated by an analytical 
continuation
from the right into the left half-plane, keeping the integrand in
(\ref{igral1}) real. This analytical continuation can be achieved by a 
joint rotation in the complex $g$-plane and of the integration 
contour in the $\vec{r}=(x,y)$-plane.
The convergence of the integral is maintained by the substitution 
$g \rightarrow g \exp(i\theta)$ 
and $\vec{r} \rightarrow \vec{r} \exp(-i\theta/4)$, where
$\theta$ is the rotation angle in the complex $g$-plane. Let us assume that
the function $Z(g,\delta)$ is analytic in the $g$-plane, with a cut along
the negative $g$-axis, and a discontinuity for $g<0$. Then the rotation in the
complex $g$-plane by an angle $\theta=\pm \pi$ yields on the lower lip of the 
cut:
\begin{equation}
\label{igral15}
Z_{\pm}(-|g|,\delta)=Z\left(|g|e^{\pm i\pi},\delta \right) \, .
\end{equation}  
The corresponding rotated integration contours 
$\left(\Gamma_{\mp}\right)$ are drawn in Figure\ \ref{grplane}.
The discontinuity across the cut is given by
\begin{equation}
\label{igral16}
{\rm Disc}\, Z=\int_0^{2\pi}\frac{d\varphi}{2\pi} \int_{\Gamma}rdr\exp\left[
-\frac{r^2}{2}-G(g,\delta,\varphi)\frac{r^4}{4}\right]\, ,
\end{equation}
where the combined contour $\Gamma=\Gamma_{+}-\Gamma_{-}$ runs for $r>0$ 
entirely through the right half plane.

\begin{figure}[p]
\unitlength1cm
\begin{center}
\begin{picture}(5,4)
\put(0,0){\epsfxsize=5cm \epsfbox{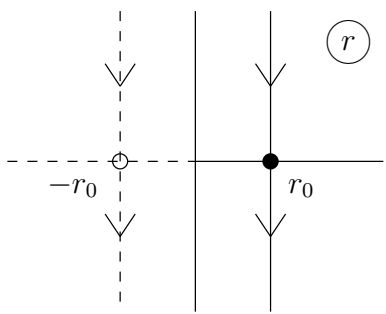}} 
\put(0.5,1.6){$-r_0$}
\put(3.7,1.6){$r_0$}
\put(4.5,3.6){\circle{0.6}} 
\put(4.4,3.5){$r$} 
\end{picture}
\end{center}
\caption{\label{intcont} Deformation of the contours of integration to
make them pass through the saddle point $r_0$.} 
\end{figure}

In a perturbatively expansion in powers of $\delta$, the discontinuity
can be computed from an expansion around the saddle point
\begin{equation}
\label{igral17}
r_0=\sqrt{\frac{1}{|g|}} 
\end{equation} 
of the isotropic case $\delta = 0$.
Since $r>0$, only the positive square root contributes, the negative one is
automatically taken into account by the integration over the angle $\varphi$.
Now, the contour of integration $\Gamma$ in the right half-plane can be
deformed to run vertically across the 
saddle point (see Figure \ref{intcont}), i.\ e.\ ,
we can integrate along a straight line:
\begin{equation}
r=\sqrt{\frac{1}{|g|}}-i\xi \, .
\end{equation}
The exponent in (\ref{igral16}) plays the role of an action, and the
deviations $\xi$ may be considered as fluctuations around the extremal
solution. The angle $\varphi$ is analogous to a collective coordinate
along the motion of the instanton in the isotropic limit.
Expanding the action up to the second order in $\xi$ around the 
extremum of the isotropic action, we obtain 
\begin{eqnarray}
\label{igral18}
{\rm Disc} \, Z\!\!\!&=&\!\!\!-\frac{i}{2\pi}\left(\frac{1}{|g|}\right)^{1/2}
\int_{-\infty}^{+\infty}\!\!\int_0^{2\pi}d\xi d\varphi
\nonumber \\
      & &\times \exp\left[-\frac{1}{4|g|}-
         \frac{\delta}{8|g|}\sin^2\left(2\varphi\right)-
         {\xi}^2+{\cal} O\left(\frac{\delta}{\sqrt{|g|}}\right)\right]\, .
\end{eqnarray}
Integrating out the fluctuations $\xi$ and the collective coordinate
$\varphi$, and using the equation
\begin{equation}
{\rm Im} \, Z=\frac{1}{2i} {\rm Disc} \, Z
\end{equation}
we obtain the following imaginary part for $Z$:
\begin{equation}
\label{igral19}
{\rm Im}\, Z=-\sum_{n=0}^{\infty}(-1)^n {\delta}^n \frac{\Gamma(n+\frac{1}{2})}{
2^n \Gamma(n+1)^2}\left(\frac{1}{4|g|}\right)^{n+1/2}\exp\left(-\frac{1}{
4|g|}\right)\left[1+{\cal}O(g)\right] \, .
\end{equation}
Each power ${\delta}^n$ has its own $n$-dependent imaginary part. Given such
an expansion, the large-order estimates for the coefficients 
$Z_{kn}$ (with $k\ll n$) follows from a dispersion relation in $g$ 
(see for example Eq.\ (6) in Ref. \cite{Jan})
\begin{equation}
\label{igral20}
Z_{kn}=\frac{1}{\pi}\int_{-\infty}^0 dg\frac{{\rm Im}\, Z^{(n)}(g+i0)}{g^{k+1}} \, ,
\end{equation} 
where $Z^{(n)}(g)$ is the coefficient of ${\delta}^n$.
In general if a real analytic function $F(g)$ has on top of the cut 
along $g\in (-\infty,0)$ an imaginary part
\begin{equation}
\label{igral21}
{\rm Im}\, F(g+i0)=-\pi\gamma\left(\frac{1}{\sigma |g|}\right)^{\beta+1}\exp\left(
-\frac{1}{\sigma |g|}\right)\left[1+{\cal O}(g)\right]\, ,
\end{equation}
then a dispersion relation of the form (\ref{igral20}) leads to the 
asymptotic behavior
\begin{equation}
\label{igral22}
F_k=\gamma (-1)^k {\sigma}^k k^{\beta} k! \left[1+{\cal O}(1/k)\right] \, .
\end{equation}
With $\sigma =4$ and $\beta=n-1/2$, we obtain again the result (\ref{igral14}).

Thus, the steepest descent method using the isotropic saddle point is a
perfect tool for calculating the large-orde behavior of the expansion
coefficients $Z_{kn}$ in the expansion (\ref{igral12}).
A great advantage of this method with respect to the exact calculation
(\ref{igral13}) is the fact that it can be generalized to quantum mechanics
and field theory where exact calculations would be unpossible.

Before applying the resummation algorithm of the previous section we have to
study the strong-coupling behavior, i.\ e., the limit of large $g$. This can
simply be done by rescaling the integral (\ref{igral2})
\begin{equation}
\label{igral23}
Z=\int_0^{\infty}dy \int_0^{2\pi}\frac{d\varphi}{2\pi}
G(g,\delta,\varphi)^{-1/2}\exp\left(-\frac{y}{\sqrt{G(g,\delta,\varphi)}}-
y^2\right)
\end{equation}
with $G$ from (\ref{igral3}) and $y=\rho\sqrt{G}$.
Taking the limit of large $g$ (i.\ e.\ large $G$) and integrating out
the angle $\varphi$ we find
\begin{equation}
\label{igral24} 
Z(g,\delta))\stackrel{g \rightarrow \infty}{\longrightarrow}\,
\kappa(\delta) g^{-1/2}
\end{equation}  
with 
\begin{equation}
\kappa(\delta)=\frac{{\pi}^{1/2}}{2}\sum_{n=0}^{\infty}\frac{
(2n)!^2}{(n!)^4 2^{5n}} {\delta}^n \, .
\end{equation}
Now, a resummation of the $g$-series in (\ref{igral12}) yields a
generalization of (\ref{res5}): 
\begin{equation}
\label{igral25}
Z^{(N)}(g,\delta) \equiv \sum_{n=0}^N 
\left(\sum_{p=n}^N a_{pn}\, I_{pn}(g)\right)
{\delta}^n
\end{equation}
with the complete set of Borel summable functions
\begin{eqnarray}
\label{igral26}
\lefteqn{I_{pn}(g)=} \nonumber \\
& &\left(\frac{4}{\sigma g}\right)^{b_0(n)+1}\int_0^1 dw 
   \frac{(1+w) w^{b_0(n)+p}}{\Gamma\left[b_0(n)
   +1\right](1-w)^{2b_0(n)+2\alpha+3}}
   \exp\left[-\frac{4w}{(1-w)^2 \sigma g}\right]\, , \nonumber \\
\end{eqnarray} 
and the coefficients 
\begin{equation}
\label{igral27}
a_{pn}=\sum_{k=n}^p \frac{Z_{kn}}{(b_0(n)+1)_k} 
\left(\frac{4}{\sigma}\right)^k
\left(\begin{array}{c}
        p+k-1-2\alpha \\ p-k
       \end{array} \right)\, ,
\end{equation}  
where the perturbation coefficients $Z_{kn}$ are given by (\ref{igral13}).
The parameters $b_0(n)$, $\sigma$ and $\alpha$ follow from the
large-order behavior (\ref{igral14}) and the strong-coupling expansion
(\ref{igral24}), respectively:
\begin{eqnarray}
\label{igral28}
b_0(n)\!\!\!&=&\!\!\!n+1 \nonumber \\
\sigma\!\!\!&=&\!\!\!4 \nonumber \\
\alpha\!\!\!&=&\!\!\!-\frac{1}{2} \, .
\end{eqnarray}
From (\ref{igral27}) it is possible to derive the following closed 
formula for the coefficients $a_{pn}$:
\begin{equation}
\label{igral29}
a_{pn}=\frac{2^n}{\pi}\frac{\Gamma\left(n+\frac{1}{2}\right)^2\Gamma(n+2)}{
\Gamma(2n+2)} (-1)^p \frac{(-p)_n (p-n)!}{p!\, n!}
\left(\begin{array}{c}
        2\alpha+1 \\ p-n
       \end{array} \right)\, ,
\end{equation}
Inserting the exact strong-coupling parameter $\alpha=-1/2$, we obtain
\begin{equation}
\label{igral30}
a_{pn}=\left\{
\begin{array}{cl}
\displaystyle{
\frac{1}{8^n}\frac{(n+1)}{(2n+1)}\frac{(2n)!}{(n!)^2} }&\mbox{for $p=n$}\, , \\
                                                   0 & \mbox{else} \, . \\
\end{array}
\right.          
\end{equation}
Thus we have a general result that the approximants $\sum_{p=n}^N a_{pn}\,
I_{pn}(g)$ in (\ref{igral25}) posses no terms with $p>n$, where $n$ is the
power of $\delta$. In such a way the $n$-dependent functions of $g$
associated to each ${\delta}^n$ are recovered exactly. 

\begin{figure}[p]
\unitlength1cm
\vspace{-8ex}
\begin{minipage}[t]{13.2cm} 
\epsfxsize=13cm
\hfill
\epsfbox{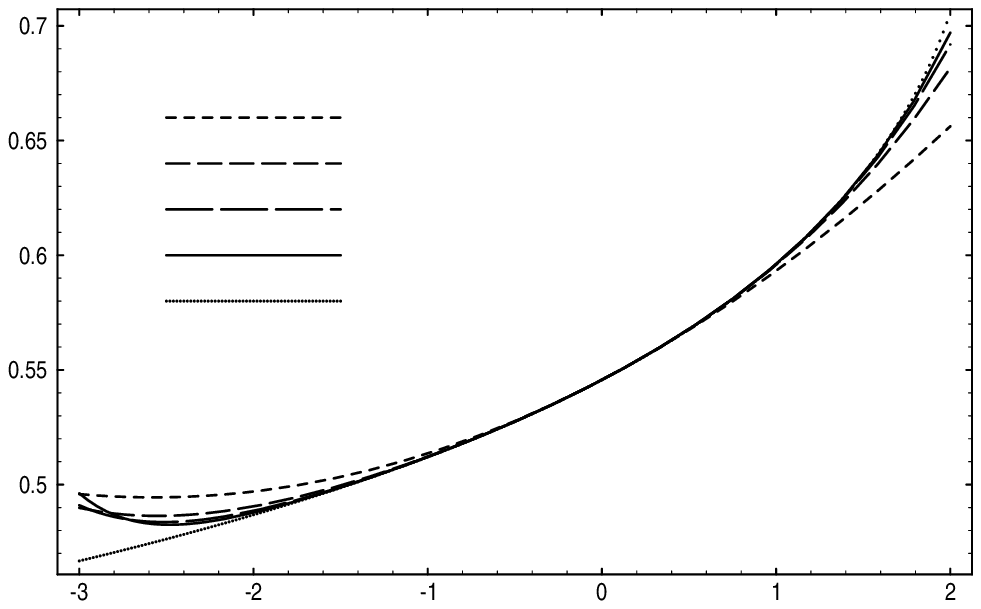} \par
\begin{picture}(13,2)
\put(0,10){{\it Z}}
\put(5.5,11.2){{\it N} = 2}
\put(5.5,10.6){{\it N} = 4}
\put(5.5,10.0){{\it N} = 6}
\put(5.5,9.4){{\it N} = 8}
\put(5.5,8.8){exact}
\put(7.0,4.25){$\delta$}
\end{picture}
\end{minipage}
\vspace{-20ex} 
\caption{\label{inp25} Partition function {\it Z} of the 
simple integral model as a function of the
anisotropy \mbox{parameter $\delta$} with the coupling constant $g/4=0.25$.
Comparison is made between the precise numerical result and the resummed 
perturbation series $Z^{(N)}$ [see Eq. (\ref{igral25})] 
for various order {\it N}.}
\end{figure}

\begin{figure}[p]
\unitlength1cm
\vspace{-8ex}
\begin{minipage}[t]{13.2cm} 
\epsfxsize=13cm
\hfill
\epsfbox{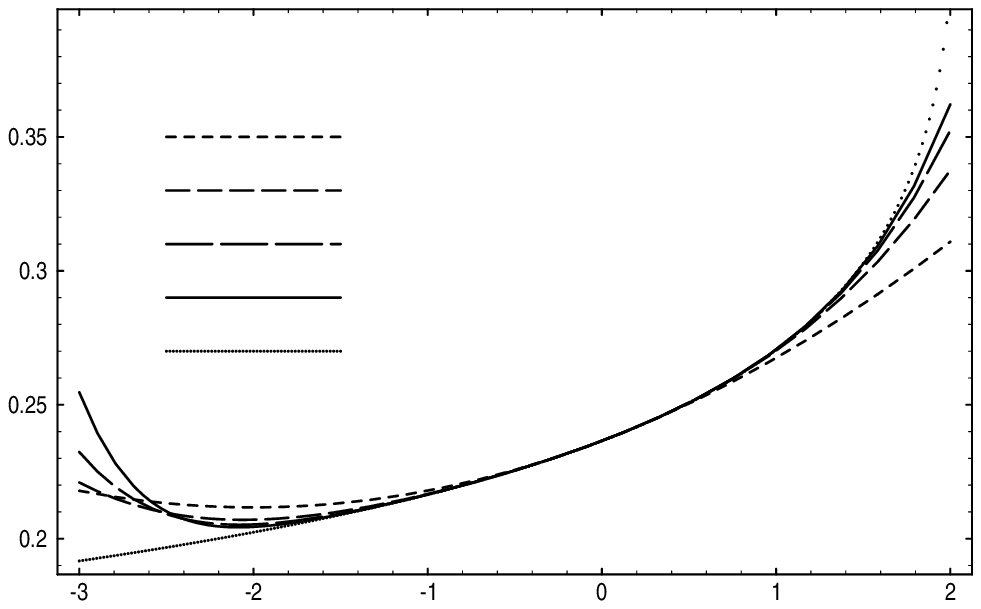} \par
\begin{picture}(13,2)
\put(0,10){{\it Z}}
\put(5.5,10.95){{\it N} = 2}
\put(5.5,10.25){{\it N} = 4}
\put(5.5,9.55){{\it N} = 6}
\put(5.5,8.85){{\it N} = 8}
\put(5.5,8.15){exact}
\put(7.0,4.25){$\delta$}
\end{picture}
\end{minipage}
\vspace{-20ex} 
\caption{\label{in2p5} The same functions as in Fig. \ref{inp25}, but with
$g/4$ equal to $2.5$.}
\end{figure}

The approximation $Z^{(N)}(g,\delta)$ may then be compared with the 
numerically calculated integral (\ref{igral4}). 
In Figures \ref{inp25} and \ref{in2p5} we have shown the result for 
various $N$ and coupling 
constants $g/4$. 
\section{Quantum mechanics}
After the integral model, the simplest nontrivial example for our 
approximation method is a $\phi^4$-theory in one spacetime dimension 
with a cubic anisotropy, which is equivalent to the quantum mechanics of an 
anisotropic anharmonic oscillator.
\subsection{Recursions relations for the ground-state perturbation 
coefficients} 
Consider an anharmonic oscillator with cubic anisotropy and a
Hamilton operator
\begin{equation}
\label{quant1} 
H=-\frac{1}{2}\left(\frac{\partial^2}{\partial x^2}+
\frac{\partial^2}{\partial y^2}
\right)+\frac{\omega^2}{2} \left(x^2+y^2\right)+\frac{g}{4}\left[x^4+
2(1-\delta)x^2 y^2+y^4 \right] \, .
\end{equation}  
Introducing reduced variables by a rescaling
\begin{eqnarray}
\label{quant2}
x &\rightarrow &\sqrt{\frac{1}{\omega}} x \quad , \quad
y \rightarrow \sqrt{\frac{1}{\omega}} y \nonumber \\
g &\rightarrow & {\omega}^3 g \quad , \quad
E^{(0)} \rightarrow  \omega E^{(0)} \, ,
\end{eqnarray}
yields the dimensionless time-independent Schr\"odinger equation 
\begin{equation}
\label{quant3}
\left[-\frac{1}{2}\left(\frac{\partial^2}{\partial x^2}+
\frac{\partial^2}{\partial y^2}\right)+\frac{1}{2} r^2 
+\frac{g}{4}\left(r^4-2\delta x^2 y^2\right) \right]
\Psi_n(x,y)=E^{(n)} \Psi_n(x,y) \, . 
\end{equation}
with the associated boundary condition
\begin{equation}
\label{quant4}
|\Psi_n(x,y)| \longrightarrow 0 \quad , \quad
r= \sqrt{x^2+y^2} \longrightarrow \infty \,\, .
\end{equation}
The boundery condition  selects only the discrete energy eigenvalues
$E^{(n)}$. We now consider the ground-state energy $E^{(0)}=E$, whose
perturbation expansion has the form
\begin{equation}
\label{quant5}
E=\sum_{m=0}^{\infty} \sum_{l=m}^{\infty} \left(\frac{g}{4}\right)^l
(2\delta)^m E_{lm}
\end{equation} 
with the unperturbed ground-state energy $E_{00}=1$.
In the following, we refer to (\ref{quant5}) as the Rayleigh-Schr\"odinger
series and the $E_{lm}$ as a Rayleigh-Schr\"odinger coefficient.

In general, the ground-state energy is available from the sum of all
connected Feynman diagrams having no externel legs. For an efficient
computation of the Rayleigh-Schr\"odinger coefficients at large orders we
shall derive recursions for the $E_{lm}$ following a method introduced 
by Bender and Wu \cite{BeWu}. 
In this way we obtain a difference equation generating 
the Rayleigh-Schr\"odinger coefficients. 

Separating out the unperturbed ground-state wave function,
$\Psi_0(x,y)=\exp[-(x^2+y^2)/2]$, we substitute
\begin{equation}
\label{quant6}
\Psi(x,y)=\sum_{n=0}^{\infty} \sum_{k=n}^{\infty}
\left(-\frac{g}{4}\right)^k (2\delta)^n
\exp[-(x^2+y^2)/2] \Phi_{kn}(x,y) \, ,
\end{equation} 
where $\Phi_{kn}(x,y)$ is polynomial in $x,y$ with $\Phi_{00}=1$.
Inserting the perturbation expansions (\ref{quant5}), (\ref{quant6}) into
the differential equation (\ref{quant3}), and collecting powers of $g$ and
$\delta$, we find
\begin{eqnarray}
\label{quant7}
-\frac{1}{2}\left(\frac{\partial^2}{\partial x^2}+
\frac{\partial^2}{\partial y^2}\right)\Phi_{kn}+
\left(x \frac{\partial}{\partial x}+
y \frac{\partial}{\partial y} \right)\Phi_{kn}-
r^4 \Phi_{k-1,n}+x^2 y^2 \Phi_{k-1,n-1}&=& \nonumber \\
\sum_{l=1}^k (-1)^l E_{l0} \Phi_{k-l,n}+
\sum_{m=1}^n \sum_{l=m}^k
(-1)^l E_{lm} \Phi_{k-l,n-m}& &\!\!\!\!\!\!\!\!\!\!. 
\end{eqnarray}
Finally, the ansatz
\begin{equation}
\label{quant8}
\Phi_{kn}(x,y)=\sum_{i,j=0}^{2k-n} A_{ij}^{kn} x^{2i} y^{2j}
\end{equation}
with
\begin{equation}
\label{quant9}
A_{ij}^{kn}=0 \quad {\rm for} \quad i,j>2k-n \quad ; \quad i,j,k,n<0
\quad ; \quad k<n
\end{equation}
gives the desired difference equation
\begin{eqnarray}
\label{quant10}
2(i+j)A_{ij}^{kn}&=&(2i+1)(i+1)A_{i+1,j}^{kn}+(2j+1)(j+1)A_{i,j+1}^{kn}
\nonumber \\
& &\!\!\! +A_{i-2,j}^{k-1,n}+A_{i,j-2}^{k-1,n} 
+2A_{i-1,j-1}^{k-1,n}-A_{i-1,j-1}^{k-1,n-1}
\nonumber \\ 
& &\!\!\!-\sum_{l=1}^k\left(A_{10}^{l0}+A_{01}^{l0}\right)A_{ij}^{k-l,n}-
\sum_{m=1}^n\sum_{l=m}^k \left(A_{10}^{lm}+A_{01}^{lm}
\right)A_{ij}^{k-l,n-m}\, . 
\nonumber \\
\end{eqnarray}
The $A_{ij}^{kn}$ yield the desired Rayleigh-Schr\"odinger coefficients
$E_{kn}$ via the simple formula
\begin{equation}
\label{quant11}
E_{kn}=-(-1)^k \left(A_{10}^{kn}+A_{01}^{kn}\right)\, .
\end{equation}
These can be determined recursively
via (\ref{quant10}). The recursion must be initialized with
\begin{equation}
\label{quant12}
A_{00}^{kn}={\delta}_{k0}\, {\delta}_{n0} \, ,
\end{equation}
and solved at increasing $k=0,1,2,\ldots$; $n=0,1,2,\ldots,k$
and, for each set $k$ and $n$, 
with decreasing $i=2k-n,\ldots,0$; $j=2k-n,\ldots,0$ (omitting $i$=$j$=0). 
The procedure is most easily performed with the help of an 
algebraic computer program. 
The list of the first Rayleigh-Schr\"odinger coefficients
up to $k=12$ $(n=0,\ldots,k)$ is given Table \ref{rscoeff}.
\begin{table}[p]
\begin{minipage}[t]{5cm} \footnotesize
\begin{tabular}[b]{|c|c|l|}  \hline \hline 
k & n & \hspace{8ex} $E_{kn}$ \\ \hline 
0 & 0 & 1 \\ [1ex]
1 & 0 & 2  \\
1 & 1 & -1/4 \\ [1ex]
2 & 0 & -9 \\
2 & 1 & 9/4 \\
2 & 2 & -3/16 \\  [1ex]
3 & 0 & 89 \\
3 & 1 & -267/8 \\
3 & 2 & 177/32 \\
3 & 3 & -11/32 \\ [1ex]
4 & 0 & -5013/4 \\
4 & 1 & 5013/8 \\
4 & 2 & -9943/64 \\
4 & 3 & 2465/128 \\
4 & 4 & -973/1024 \\ [1ex]
5 & 0 & 88251/4 \\
5 & 1 & -441255/32 \\
5 & 2 & 874757/192 \\
5 & 3 & -216751/256 \\
5 & 4 & 171049/2048 \\
5 & 5 & -20987/6144 \\ [1ex]
6 & 0 & -3662169/8 \\
6 & 1 & 10986507/32 \\
6 & 2 & -327063703/2304 \\
6 & 3 & 81133049/2304 \\
6 & 4 & -64093757/12288 \\
6 & 5 & 31487347/73728 \\
6 & 6 & -4401593/294912 \\ [1ex]
7 & 0 & 86716929/8 \\
7 & 1 & -607018503/64 \\
7 & 2 & 32603176343/6912 \\
7 & 3 & -40534191905/27648 \\
7 & 4 & 32089547489/110592 \\
7 & 5 & -15794879119/442368 \\
7 & 6 & 4423646695/1769472 \\
7 & 7 & -135064261/1769472 \\ 
\end{tabular}
\end{minipage} 
\hfill
\begin{minipage}[t]{7.7cm} \footnotesize
\begin{tabular}[b]{|c|c|l|}
8 & 0 & -18380724429/64 \\
8 & 1 & 18380724429/64 \\
8 & 2 & -6932983533833/41472 \\
8 & 3 & 216189163547/3456 \\
8 & 4 & -6865860756773/442368 \\
8 & 5 & 847050762955/331776 \\
8 & 6 & -951145969207/3538944 \\
8 & 7 & 116411434099/7077888 \\
8 & 8 & -151575359341/339738624 \\ [1ex]
9 & 0 & 537798950495/64 \\
9 & 1 & -4840190554455/512 \\
9 & 2 & 785448510795415/124416 \\
9 & 3 & -172109470699495/62208 \\
9 & 4 & 5484677894663731/6635520 \\
9 & 5 & -2714832036203789/15925248 \\
9 & 6 & 1910496739715441/79626240 \\
9 & 7 & -468820318449871/212336640 \\
9 & 8 & 1223678377567247/10192158720 \\
9 & 9 & -29878788733243/10192158720 \\ [1ex]
10 & 0 & -34427971992123/128 \\
10 & 1 & 172139859960615/512 \\
10 & 2 & -757445337006448801/2985984 \\
10 & 3 & 95243865818145949/746496 \\
10 & 4 & -26657139955813121627/597196800 \\
10 & 5 & 6619289843855618939/597196800 \\
10 & 6 & -18688108386867852767/9555148800 \\
10 & 7 & 287396519063579707/1194393600 \\
10 & 8 & -6016110774357344761/305764761600 \\
10 & 9 & 588950135128273907/611529523200 \\
10 & 10 & -52319976745196951/2446118092800 \\ 
\end{tabular}
\end{minipage}
\end{table}
\begin{table}[p]  \footnotesize
\begin{center}
\begin{tabular}[t]{|c|c|l|} 
11 & 0 & 1196938085820951/128 \\
11 & 1 & -13166318944030461/1024 \\
11 & 2 & 484953641311740249799/44789760 \\
11 & 3 & -122498739278392549037/19906560 \\
11 & 4 & 273164737489274832749/110592000 \\
11 & 5 & -8576021167229490768493/11943936000 \\
11 & 6 & 485748876580259709683/3185049600 \\
11 & 7 & -11098068163230668731/471859200 \\
11 & 8 & 786108601809348099491/305764761600 \\
11 & 9 & -385711575108432009551/2038431744000 \\
11 & 10 & 7631665291905150913/905969664000 \\
11 & 11 & -12593952190067271863/73383542784000 \\ [1ex] 
12 & 0 & -179761724871375777/512 \\
12 & 1 & 539285174614127331/1024 \\
12 & 2 & -658487704407131831592119/1343692800 \\
12 & 3 & 83537029566207575386361/268738560 \\
12 & 4 & -1870114571495468628478319/13271040000 \\
12 & 5 & 4208267075207850881725247/89579520000 \\
12 & 6 & -16737064308714173333404777/1433272320000 \\
12 & 7 & 38348532616813953055927/17694720000 \\
12 & 8 & -27234664511494120875149389/91729428480000 \\
12 & 9 & 1339555446357501564974269/45864714240000 \\
12 & 10 & -53129831724844951147579/27179089920000 \\
12 & 11 & 70291727826145874647867/880602513408000 \\
12 & 12 & -52920213881686076606297/35224100536320000 \\ \hline \hline
\end{tabular}
\end{center}
\caption{\label{rscoeff} Coefficients $E_{kn}$ in
the perturbation series (\ref{quant5}) for the ground-state energy
up to $k = 12$ ($n = 0$,\dots,$k$).}
\end{table}
\subsection{Large-order coefficients and resummation}
Working with Langer's formulation \cite{Lang} (which is related to 
Lipatov's \cite{Lipat} by a dispersion relation) and making use of known 
results for the isotropic anharmonic oscillator, we shall derive the 
large-order behavior of perturbation expansion for the ground-state energy.

The method is based on the path-integral representation of the quantum
partition function
\begin{equation}
\label{quant21}
Z=\int {\cal D}x{\cal D}y \exp[-A(x,y)]
\stackrel{\beta \rightarrow \infty}{\longrightarrow}\, \exp(-\beta E)
\end{equation}  
where
\begin{equation}
\label{quant22}
A=\int_{-{\beta}/2}^{+{\beta}/2}d\tau \left\{\frac{1}{2}\left(
\dot{x}^2+\dot{y}^2\right)+\frac{1}{2}\left(x^2+y^2\right)+\frac{g}{4}\left[
x^4+2(1-\delta)x^2y^2+y^4\right] \right\}
\end{equation}
is the Euclidean action corresponding to the Hamiltonian (\ref{quant1}).
For $g>0$, the system is stable and $Z$ is real. On the other hand, if the
coupling constant $g$ is negative the system becomes unstable and $Z$
develops an exponentially small imaginary part related to the decay-rate
$\Gamma$ of the ground-state resonance. The imaginary part of the
ground-state energy may be obtained by taking the large $\beta$ limit in
(\ref{quant21}),
\begin{equation}
\label{quant23}
{\rm Im}\, E=\frac{1}{2} \Gamma=
-\frac{1}{\beta}\frac{{\rm Im}\, Z}{{\rm Re}\, Z} \quad ,
\quad \beta \rightarrow \infty \, .
\end{equation}
In the above equation the fact was used that ${\rm Im}\, Z \propto
\exp(-\beta) \exp\left[-1/(\sigma |g|)\right]$ is much smaller than
${\rm Re} \, Z=\exp\{-\beta[1+{\cal O}(g)]\}$. 
For small $g<0$, this imaginary part
can be computed perturbatively in the anisotropic parameter $\delta$ by the
expansion around the isotropic instanton solution $r_c(\tau)$:
\begin{equation}
\label{quant24}
\left(
\begin{array}{c}
  x \\ y
\end{array} \right)=
\left(
\begin{array}{c}
  \cos\varphi \\ \sin\varphi
\end{array} \right) (r_c+\xi)+
\left(
\begin{array}{r}
 -\sin\varphi \\ \cos\varphi
\end{array} 
\right)\eta \,\, , \, \, r_c=\sqrt{\frac{2}{|g|}}\frac{1}{\cosh(
\tau-\tau_0)} \, .
\end{equation}
For simplicity, we shall set $\tau_0=0$ in the sequel. In (\ref{quant24})
we have separated out the rotation angle $\varphi$ of the isotropic
instanton in the $(x,y)$-plane and $\xi$, $\eta$ are the degrees of freedom,
orthogonal to this rotation. 
Inserting the expansion (\ref{quant24}) into the action (\ref{quant22}) we
obtain the expression
\begin{eqnarray}
\label{quant25}
A&=&\frac{4}{3|g|}+\frac{\delta}{|g|}\frac{2\sin^2(2\varphi)}{3} 
\nonumber \\
 & &+\frac{1}{2}\int d\tau \left[\xi \left(-\frac{d^2}{d\tau^2}+1-\frac{6}{
\cosh^2\tau}\right)\xi+\eta\left(-\frac{d^2}{d\tau^2}+1-\frac{2}{\cosh^2\tau}
\right)\eta\right]
\nonumber \\
 & &+{\cal O}\left(\frac{\delta}{\sqrt{|g|}} \right) \, ,
\end{eqnarray}
where we have splitted of the action into the terms responsible for the
leading contributions in an expansion of the form (\ref{igral19}) and a
remainder ${\cal O}\left({\delta}/{\sqrt{|g|}}\right)$. 
Then the $\delta$-dependence of the transversal quadratic 
fluctuations  belongs to the omitted terms.
Expanding (\ref{quant21}) in $\delta$ and integrating out the quadratic
fluctuations we obtain
\begin{equation}
\label{quant26}
Z=f_{\xi}f_{\eta}\int_0^{2\pi}d\varphi\sum_{n=0}^{\infty}
\frac{(-{\delta}/4)^n}{n!}\left[2\sin^2(2\varphi)\right]^n \left(\frac{4}{
3|g|}\right)^n \exp\left(-\frac{4}{3|g|}\right)[1+{\cal O}(g)]  
\end{equation}  
where the angle integral can be done with 
\begin{equation}
\label{quant27}
\int_0^{2\pi}d\varphi \left[2\sin^2(2\varphi)\right]^n=8^n 2
\frac{\Gamma^2\left(n+\frac{1}{2}\right)}{\Gamma(2n+1)}=
8^n 2 B\left(n+\frac{1}{2},n+\frac{1}{2}\right) \, .
\end{equation}
The contribution $f_{\xi}$ and $f_{\eta}$ from the quadratic longitudinal
and transversal fluctuations coincide with those appearing in the isotropic
oscillator problem and are therefore known. With the isotropic classical
action $A_{0c}=4/(3|g|)$ the well-known results are
\begin{eqnarray}
f_{\xi}&=&-\frac{i}{2}\sqrt{\frac{A_{0c}}{2\pi}}\beta 
\left| 
\frac{\det'(-d^2 /d{\tau}^2+1-6/{\cosh}^2{\tau})}{\det(-d^2 /d{\tau}^2+1)}
\right|^{-1/2} Z_{{\rm osc}}
\nonumber \\
      &=&-\frac{i}{2}\sqrt{\frac{A_{0c}}{2\pi}}\beta \sqrt{12}
\exp(-{\beta}/2)
\nonumber 
\end{eqnarray}
and
\begin{eqnarray}
\label{quant28}
f_{\eta}&=&\sqrt{\frac{3A_{0c}}{2\pi}}
\left[ 
\frac{\det'(-d^2 /d{\tau}^2+1-2/{\cosh}^2{\tau})}{\det(-d^2 /d{\tau}^2+1)}
\right]^{-1/2} Z_{{\rm osc}}
\nonumber \\
        &=& 2 \sqrt{\frac{3A_{0c}}{2\pi}} \exp(-{\beta}/2) 
\end{eqnarray}
where we have used the partition function of the harmonic oscillator
\begin{equation}
\label{quant29}
Z_{{\rm osc}} \equiv \det(-d^2 /d{\tau}^2+1)^{-1/2}=\frac{1}{2\sinh({\beta}/2)}
\stackrel{\beta \rightarrow \infty}{\longrightarrow}\, \exp(-{\beta}/2)
\end{equation}
to normalize the determinants. In the upper determinants, the zero eigenvalues
are excluded. This fact is recorded by the prime.

The longitudinal fluctuations $\xi$ contain a negative eigenmode, this being
responsible for the factor $-i/2$ and the absolut value sign, and a zero
eigenmode associated with the translation invariance which is spontaneously
broken by the special choice ${\tau}_0=0$. The separation of this zero
eigenmode in the framework of collective coordinates yields the factor
$\beta \sqrt{A_{0c}/(2\pi)}$ (see Chapter 17 in Ref. \cite{book}). 
Collecting the contributions of the negative
and all positive eigenmodes one obtains the remaining factor $\sqrt{12}$.

In contrast to the longitudinal case the transversal fluctuations $\eta$
do not contain any negative mode. The transversal fluctuation operator has
one zero eigenvalue due to the rotational invariance in the limit
$\delta \rightarrow 0$. The associated eigenmode is extracted from the
integration measure via the change of variables (\ref{quant24}). The
Jacobian of this coordinate treansformation can be deduced from the
isotropic system. It contributes the factor $\sqrt{3A_{0c}/(2\pi)}$. The
remaining factor $2$ results from all other modes with positive eigenvalues.

Collecting all contributions to the imaginary part of the ground-state energy
(\ref{quant23}), an cancellation of all $\beta$ dependent factors leads to:
\begin{eqnarray}
\label{quant30}
{\rm Im}\!\!\!\!\! &E& 
\stackrel{g \rightarrow 0^{-}}{\longrightarrow}\, 
\frac{2|f_{\xi}|f_{\eta}}{\beta \exp(-\beta)}  
\sum_{n=0}^{\infty} \frac{(-2\delta)^n}{n!}
B\left(n+\frac{1}{2},n+\frac{1}{2}\right)\left(\frac{4}{3|g|}\right)^n
\exp\left(-\frac{4}{3|g|}\right)
\nonumber \\
   & & =\frac{6}{\pi}\sum_{n=0}^{\infty}\frac{(-2\delta)^n}{n!}
B\left(n+\frac{1}{2},n+\frac{1}{2}\right)\left(\frac{4}{3|g|}\right)^{n+1}
\exp\left(-\frac{4}{3|g|}\right) \, .
\end{eqnarray}
Finally, by means of the dispersion relation (\ref{igral20}) we find the
corresponding large-order behavior of the coefficients in the expansion
(\ref{quant5}):
\begin{equation}
\label{quant31}
E_{kn}\stackrel{k \rightarrow \infty}{\longrightarrow}\,
-\frac{6}{\pi^2}\frac{(-2)^n}{n!}B\left(n+\frac{1}{2},n+\frac{1}{2}\right)
(-1)^k \left(\frac{3}{4}\right)^k k! k^n \, .
\end{equation}
After having derived the large-order behavior of $E_{kn}$ and the low-order
perturbation coefficients via the Bender and Wu-like recursions
(\ref{quant10}) and (\ref{quant11}), we are in the position to resumme
the $g$-series accompanying each power ${\delta}^n$ in the expansion
(\ref{quant5}). 

The remaining strong-coupling expansion follows from Symanzik scaling
\cite{Symanz}:
\begin{equation}
\label{quant32}
E(g,\delta)=\sum_{m=0}^{\infty}{\kappa}_m(\delta)g^{(1-2m)/3}
\end{equation} 
i.\ e., the power behavior in the strong-coupling limit is given by
\begin{equation}
\label{quant33}
E(g,\delta)\stackrel{g \rightarrow \infty}{\longrightarrow}\,
{\kappa}_0 (\delta) g^{1/3} \, ,
\end{equation}  
where the $1/3$ coincides with that appearing in the one-dimensional
oscillator problem. The $\delta$-dependence enters by the prefactor
$\kappa_0$.

\begin{figure}[p]
\unitlength1cm
\vspace{-8ex}
\begin{minipage}[t]{13.2cm} 
\epsfxsize=13cm
\hfill
\epsfbox{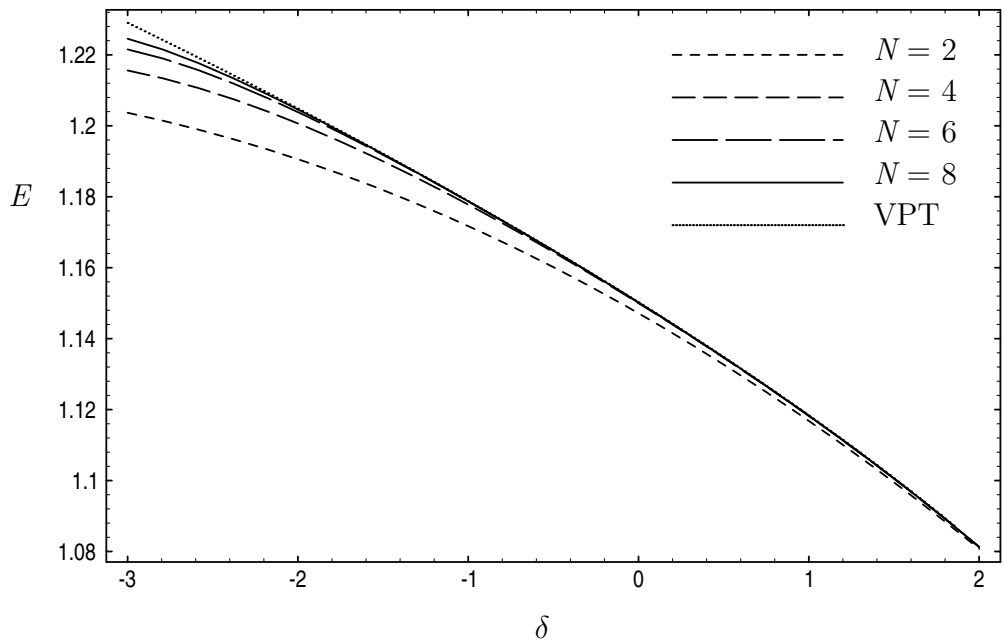} \par
\begin{picture}(13,2)
\put(0,10){{\it E}}
\put(11.5,11.95){{\it N} = 2}
\put(11.5,11.40){{\it N} = 4}
\put(11.5,10.85){{\it N} = 6}
\put(11.5,10.30){{\it N} = 8}
\put(11.5,9.75){VPT}
\put(7.0,4.25){$\delta$}
\end{picture}
\end{minipage}
\vspace{-20ex} 
\caption{\label{qmp1} Ground-state energy {\it E} of the anisotropic
anharmonic oscillator with $g/4=0.1$ as a function of the anisotropy $\delta$.
Shown are the resummed perturbation series for various order of 
approximation {\it N} and the approximation $W_5({\Omega}_5)$ (see 
the text after Eq. (\ref{quant20})) from the variational perturbation 
theory (VPT).
Differences between $W_5({\Omega}_5)$ and the exact ground-state energy
are existent only on a finer energy scale. }
\end{figure}

\begin{figure}[p]
\unitlength1cm
\vspace{-8ex}
\begin{minipage}[t]{13.2cm} 
\epsfxsize=13cm
\hfill
\epsfbox{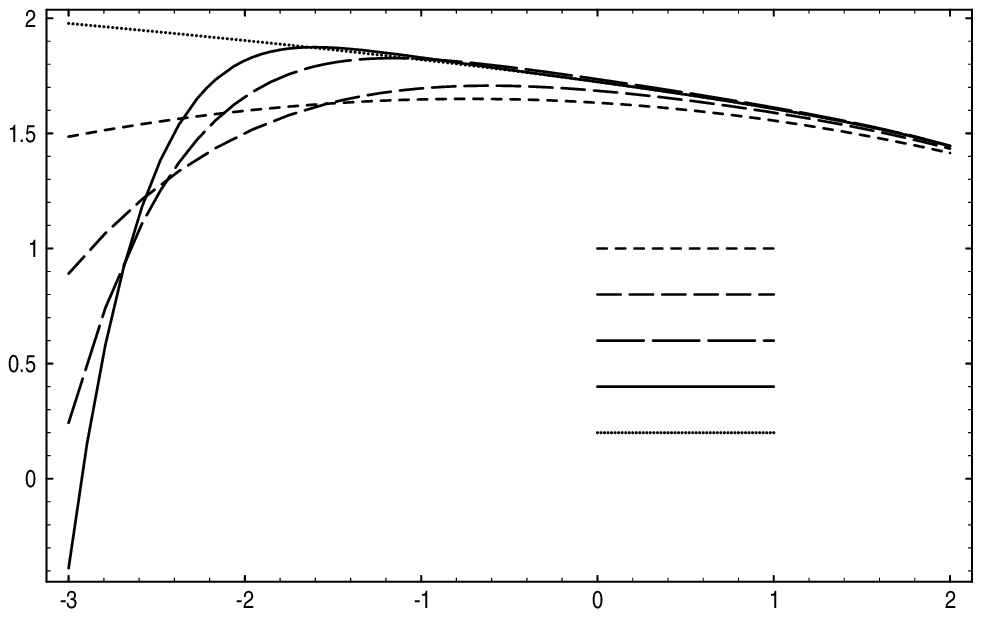} \par
\begin{picture}(13,2)
\put(0,10){{\it E}}
\put(11.0,9.6){{\it N} = 2}
\put(11.0,9.0){{\it N} = 4}
\put(11.0,8.4){{\it N} = 6}
\put(11.0,7.8){{\it N} = 8}
\put(11.0,7.2){VPT}
\put(7.0,4.25){$\delta$}
\end{picture}
\end{minipage}
\vspace{-20ex} 
\caption{\label{qm1p0} The same functions as in Fig. \ref{qmp1}, but with
$g/4=1.0$.}
\end{figure}

Combining (\ref{res15}) and the formulas (\ref{igral25}), (\ref{igral26})
and (\ref{igral27}), the resummation procedure must be worked through with
the parameters
\begin{eqnarray}
\label{quant34}
b_0(n)&=&n+\frac{3}{2}
\nonumber \\
\sigma&=&\frac{3}{4}
\nonumber \\
\alpha&=&\frac{1}{3} \, .
\end{eqnarray} 
In Figures \ref{qmp1} and \ref{qm1p0} we have plotted 
the $\delta$-dependence of the resummed
ground-state energy $E$ for two different values of the coupling constant
$g/4$ and for various orders $N$. For a
reference plot we have used the very accurate dotted curve which we have
obtained numerically from the variational perturbation theory described in the 
next subsection.
\subsection{Variational perturbation theory}
It is useful to compare the above results with those of another
recently-developed resummation procedure known as variational perturbation 
theory [for an introduction see Ref. \cite{book}, Chapter $5$]. 
Consider first the Rayleigh-Schr\"odinger expansion of the ground-state energy:
\begin{equation}
\label{quant13}
E(g,\delta)=\omega \sum_{l=0}^{\infty} \sum_{m=0}^l E_{lm}(2\delta)^m \left(
\frac{g/4}{\omega^3}\right)^l \, ,
\end{equation}
where the Rayleigh-Schr\"odinger coefficients $E_{lm}$ are obtained from the
recursion relation (\ref{quant10}) via (\ref{quant11}).

\begin{figure}[p]
\unitlength1cm
\vspace{-8ex}
\begin{minipage}[t]{13.2cm} 
\epsfxsize=13cm
\hfill
\epsfbox{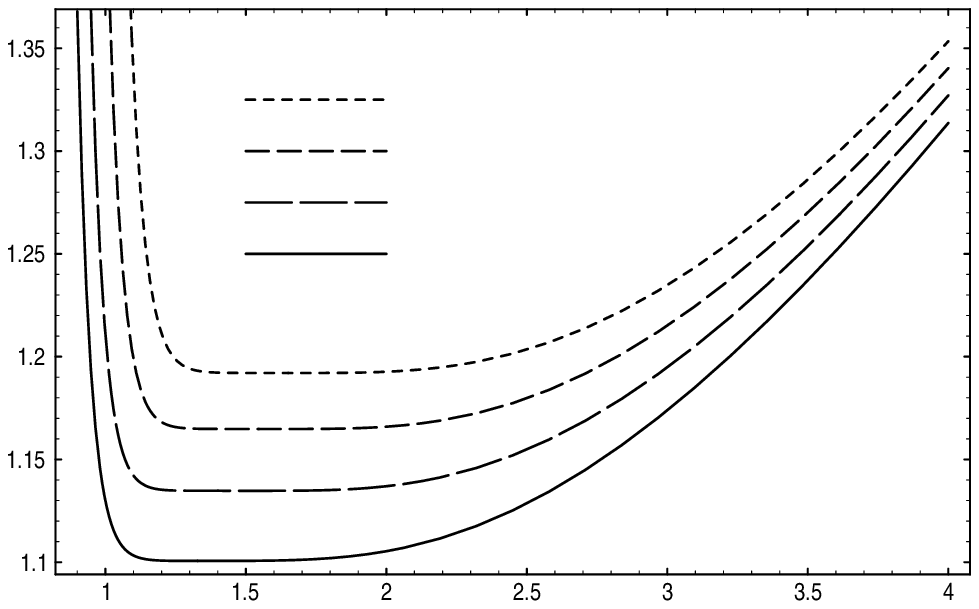} \par
\begin{picture}(13,2)
\put(0,10){{\it $W_5$}}
\put(6.1,9.4){$\delta$ = 1.5}
\put(6.1,10.1){$\delta$ = 0.5}
\put(6.1,10.8){$\delta$ = --0.5}
\put(6.1,11.5){$\delta$ = --1.5}
\put(7.0,4.25){$\Omega$}
\end{picture}
\end{minipage}
\vspace{-20ex} 
\caption{\label{W5} The function $W_5$ at constant 
coupling strength $g/4=0.1$ for various anisotropy parameters $\delta$.}
\end{figure}

\begin{figure}[p]
\unitlength1cm
\vspace{-8ex}
\begin{minipage}[t]{13.2cm} 
\epsfxsize=13cm
\hfill
\epsfbox{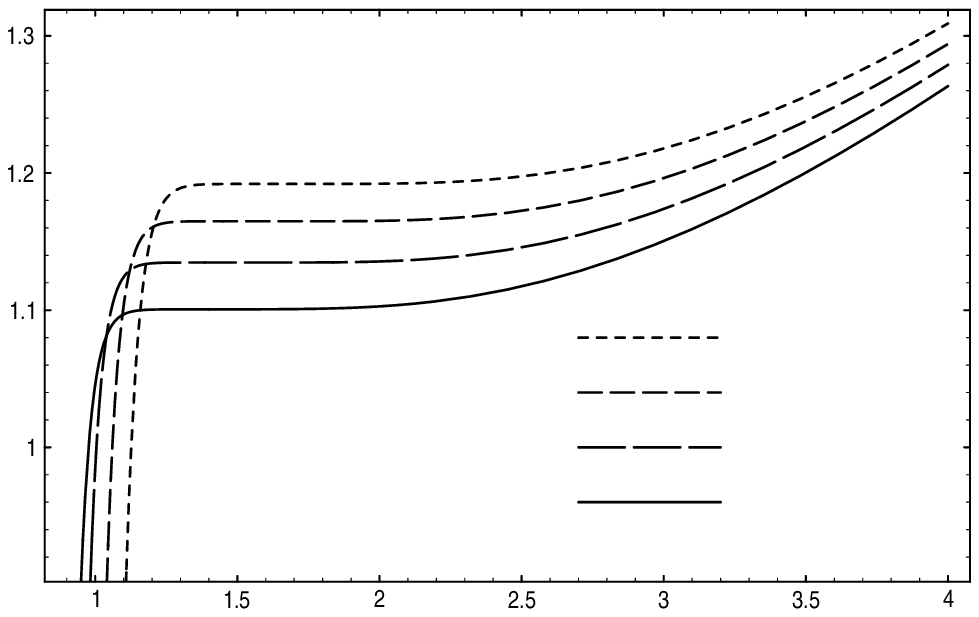} \par
\begin{picture}(13,2)
\put(0,10){{\it $W_6$}}
\put(10.4,6.20){$\delta$ = 1.5}
\put(10.4,6.95){$\delta$ = 0.5}
\put(10.4,7.70){$\delta$ = --0.5}
\put(10.4,8.45){$\delta$ = --1.5}
\put(7.0,4.25){$\Omega$}
\end{picture}
\end{minipage}
\vspace{-20ex} 
\caption{\label{W6} The function $W_6$ for $g/4=0.1$ and various $\delta$.}
\end{figure}

The variation is done as follows: 
First, the potential is separated into an arbitrary harmonic term and a 
remainder:
\begin{equation}
\label{quant14}
\frac{\omega^2}{2}\left(x^2+y^2\right)=\frac{\Omega^2}{2}\left(x^2+y^2\right)+
\frac{\omega^2-\Omega^2}{2}\left(x^2+y^2\right) \, .
\end{equation}
In contrast to ordinary perturbation theory, an interacting potential
$V_{{\rm int}}$ is defined by
\begin{equation}
\label{quant15}
V(x,y)=\frac{\Omega^2}{2}\left(x^2+y^2\right)+V_{{\rm int}}(x,y)
\end{equation}
and setting
\begin{equation}
\label{quant16}
V_{{\rm int}}(x,y)=\frac{g}{4}\left(\rho r^2+r^4-2\delta x^2 y^2\right)
\quad ; \quad \rho=\frac{2}{g}\left(\omega^2-\Omega^2\right) \, .
\end{equation}
A perturbation expansion is now found in powers of $g$ at fixed $\rho$
and $\delta$:
\begin{equation}
\label{quant17}
E_k(g,\delta,\rho)=\Omega \sum_{l=0}^k \varepsilon_l(\rho,\delta)\left(
\frac{g/4}{\Omega^3}\right)^l \, .
\end{equation}
The calculation of the new coefficients $\varepsilon_l(\rho,\delta)$ up to
a specific order $k$ does not require much additional work,
since they are easily obtained from the ordinary perturbation 
series (\ref{quant13}). 
We simply replace $\omega$ by the identical expression
\begin{equation}
\label{quant18}
\omega=\sqrt{\Omega^2+\omega^2-\Omega^2}=\sqrt{\Omega^2+g\rho/2} \, ,
\end{equation}
reexpand $E(g,\delta)$ in powers of $g$, and truncate the series after an
order $l>k$. This yields the reexpansion coefficients
\begin{equation}
\label{quant19}
\varepsilon_l(\rho,\delta)=\sum_{j=0}^l \sum_{n=0}^j E_{jn} (2\delta)^n
\left(
\begin{array}{cc}
(1-3j)/2  \\ l-j
\end{array} 
\right) 
\left(2\rho \Omega \right)^{l-j} \, . 
\end{equation}
The truncated power series
\begin{equation}
\label{quant20}
W_k(g,\delta,\Omega):=
E_k\left[g,\delta,2\left(\omega^2-\Omega^2\right)/g\right]
\end{equation}
is certainly independent of $\Omega$ for $k$ going to infinity.
However, at any finite order it depends on $\Omega$.
The optimal value of $\Omega$ is found by calculating all
extrema and the turning points.
The smallest among these order-dependent points is used as an optimal 
trying value and is denoted by $\Omega_k(g,\delta)$. 
The associated energy $W_k[g,\delta,\Omega_k(g,\delta)]$
constitutes the desired approximation to the ground-state energy.
In Figures \ref{W5} and \ref{W6} we have plotted the $\Omega$-dependence 
of $W_{5,6}$ for various anisotropy parameters $\delta$ at 
the coupling constant $g/4=0.1$.
The shape of the curves depends little on $\delta$. 
Only in the case of odd $k$ does a minimum exist. 
For even $k$, there is no extremum and the optimal $\Omega$-value lies at 
a turning point. 

For an isotropic $g x^4$-model, the precision of the variational 
perturbation method has been illustrated
by a comparison with accurate numerical energies \cite{CoKleJa}. With
increasing $k$ the approach of $W_k$ to the exact energy is quite rapid and
its mechanism is well understood.

In Table \ref{ener0} we display the ground-state energies 
for odd $k$, which
we have obtained for the anisotropic model at various $\delta$ and $g/4$. 
\begin{table}[p]
\begin{center}
\begin{tabular}{|c|lllll|} \hline \hline
\multicolumn{6}{|c|}{\rule[-3mm]{0mm}{8mm} $g/4 = 0.1$} \\ \hline \hline
k $ \backslash \delta $ &   
                -2.5     & -1.5    & -0.5     & 0.5   &  1.5 \\ \hline
1           &   1.222923 & 1.19626 & 1.167751 & 1.137 & 1.103438 \\ 
3           &   1.217193 & 1.192062 & 1.164807 & 1.134739 & 1.100658 \\ 
5           &   1.217109 & 1.192032 & 1.164801 & 1.134734 & 1.100607 \\ 
7           &   1.217107 & 1.192033 & 1.164803 & 1.134735 & 1.100604 \\ 
9           &   1.217107 & 1.192034 & 1.16481 & 1.134736 & 1.100604 \\ 
11          &   1.217107 & 1.192035 & 1.16481 & 1.134739 & 1.100604 \\   
\hline \hline
\multicolumn{6}{|c|}{\rule[-3mm]{0mm}{8mm} $g/4 = 1.0$} \\ \hline \hline
k $ \backslash \delta $ &   
                -2.5     & -1.5    & -0.5     & 0.5   &  1.5 \\ \hline
1           &   1.969986 & 1.88556 & 1.791636 & 1.684863 & 1.559412 \\ 
3           &   1.941934 & 1.863112 & 1.773978 & 1.669261 & 1.536823 \\ 
5           &   1.941196 & 1.862803 & 1.773867 & 1.669156 & 1.535609 \\ 
7           &   1.941172 & 1.862806 & 1.773888 & 1.669172 & 1.535454 \\ 
9           &   1.941172 & 1.862815 & 1.773909 & 1.669188 & 1.535425 \\ 
11          &   1.94118 & 1.862823 & 1.773924 & 1.669199 & 1.535418 \\ \hline
\end{tabular} 
\caption{\label{ener0} Convergence of the ground-state energy 
in the variational perturbation expansion for various anisotropy
\mbox{parameters $\delta$}.}
\end{center} 
\end{table}
The convergence to fixed energy values is comparable to the case of the simple
$g x^4$-interaction. So we assume that these numbers coincide with the exact
ground-state energy values at least up to the first four digits.
\section{Summary}
With the help of a simple model integral containing a quadratic and two
quartic terms of different symmetry, we have investigated in detail the
large-order behavior of the $\delta$-dependent $g$-series in a function
$f(g,\delta)=\sum_k f_k(\delta)g^k$ for the region near the isotropic
limit $\delta \rightarrow 0$. We have shown that the
large-order behavior of $f_k(\delta)$ undergoes a crossover from
the anisotropic to an isotropic regime near
the order of perturbation theory $k_{\rm cross} \approx 1/ |{\delta}|$. 

In quantum mechanics, the extreme large-order behavior of perturbation theory
for the anisotropic regime $k|{\delta}| \gg 1$ is identical with earlier
results of BBW \cite{BaBeWu} and Janke \cite{Jan}. 
In displaying the crossover-behavior we have gone beyond these earlier works.

\begin{figure}[p]
\unitlength1cm
\vspace{-8ex}
\begin{minipage}[t]{13.2cm} 
\epsfxsize=13cm
\hfill
\epsfbox{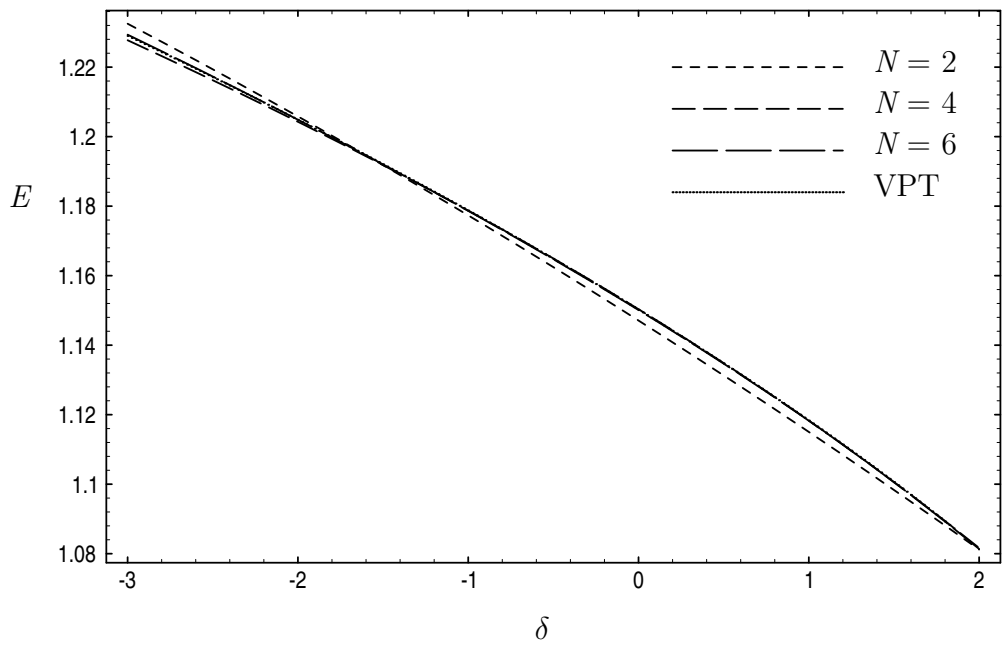} \par
\begin{picture}(13,2)
\put(0,10){{\it E}}
\put(11.5,11.80){{\it N} = 2}
\put(11.5,11.25){{\it N} = 4}
\put(11.5,10.70){{\it N} = 6}
\put(11.5,10.15){VPT}
\put(7.0,4.25){$\delta$}
\end{picture}
\end{minipage}
\vspace{-20ex} 
\caption{\label{A3p1} The same functions as in Fig. \ref{qmp1}, but with 
the large-order parameter $\sigma=3$ (explained in the text).}
\end{figure}

\begin{figure}[p]
\unitlength1cm
\vspace{-8ex}
\begin{minipage}[t]{13.2cm} 
\epsfxsize=13cm
\hfill
\epsfbox{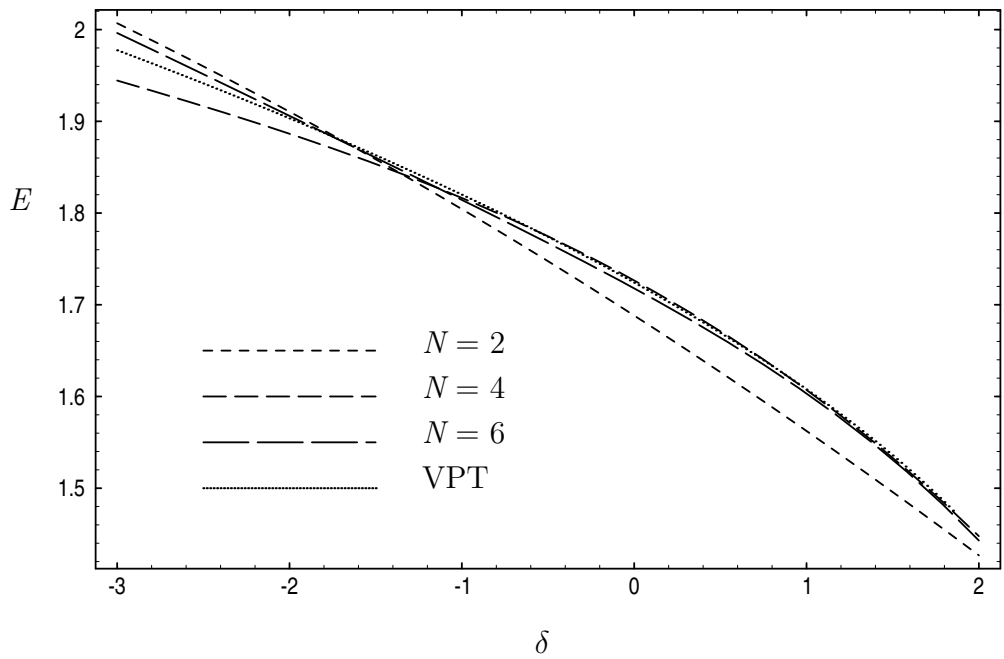} \par
\begin{picture}(13,2)
\put(0,10){{\it E}}
\put(5.5,8.1){{\it N} = 2}
\put(5.5,7.5){{\it N} = 4}
\put(5.5,6.9){{\it N} = 6}
\put(5.5,6.3){VPT}
\put(7.0,4.1){$\delta$}
\end{picture}
\end{minipage}
\vspace{-20ex} 
\caption{\label{A31} The same functions as in Fig. \ref{qm1p0}, but with 
the large-order parameter $\sigma=3$.}
\end{figure}

In particular, our resummation algorithm is shown to 
work very well in the vicinity of $\delta=0$ and for $\delta>0$, the latter
being relevant to the question of a stable cubic fixed point in field theory.
With increasing coupling
constant $g/4$, the error of the result for the ground-state energy 
becomes larger. However, for $N=6$ (this is the largest available order for 
the $\beta$-functions in quantum field theory, see Ref. \cite{KleSchu2} )
and in the wide
region $\delta \in (-0.5,2)$ and $g/4 \in (0,1)$ the error remains smaller
than $0.8 \% $. The increasing error for large negative values of $\delta$ can
intuitively be understood by comparing the first two terms in the action
(\ref{quant25}): For $\delta<0$, the ``tunneling-paths'' of extremal action
are obviously straight lines along the two diagonales in the $(x,y)$-plane
($\varphi={\pi}/4$). Along these diagonal rays, 
the basic factor $\exp[-1/(c|g|)]$
related to the decay-rate disappears for $\delta \rightarrow\, -2$, and the
ensuing expansion of (\ref{quant21}) in powers ${\delta}^n$ becomes
meaningless. An improved fit for $\delta<0$ can be obtained by choosing
larger values of the large-order parameter $\sigma$. 
In Figures \ref{A3p1} and \ref{A31} we display the result for 
$\sigma=3$ and $N=6$, where for $g/4=0.1$ the accurate and the resummed
curve coincide.

To obtain the correct description of the neighbourhood 
of the isotropic system $\delta = 0$ we have used the method developed
in the context of an anisotropic quantum field theory in \cite{KleTho}: 
By replacing the series $\sum_k f_k(\delta)g^k$ by 
$\sum_n \sum_k f_{kn}\,g^k \, {\delta}^n$ and resumming the $g$-series 
accompanying each power ${\delta}^n$, we obtain very good results
for the model integral and the ground-state energy of the anisotropic
anharmonic oscillator. In this way our results justify the earlier field 
theoretic analysis and should be useful for understanding similar problems 
in other systems. 


\newpage
\begin{thebibliography}{19}
\renewcommand{\baselinestretch}{1.0}

\bibitem{Aharon} A.\ Aharony,
               Phys.\ Rev.\ B \underline{8} (1973) 4270.

\bibitem{KetWal}   I.\ J.\ Ketley and D.\ J.\ Wallace,
                   J.\ Phys.\ A \underline{6} (1973) 1667.

\bibitem{BreGuZ1} E.\ Brezin, J.\ C.\ Le Guillou and J.\ Zinn-Justin, 
                  Phys.\ Rev.\ B \underline{10} (1974) 893. 

\bibitem{MaSo}    I.\ O.\ Mayer and A.\ I.\ Sokolov,
                  Izv.\ Akad.\ Nauk SSSR 
                  Ser.\ Fiz.\ \underline{51} (1987) 2103;
                  I.\ O.\ Mayer, A.\ I.\ Sokolov, and B.\ N.\ Shalayev, 
                  Ferroelectrics \underline{95} (1989) 93.

\bibitem{KleSchu2} H.\ Kleinert and V.\ Schulte-Frohlinde, 
                   Phys.\ Lett.\ B \underline{342} (1995) 284. 

\bibitem{BaBeWu}  T.\ Banks, C.\ M.\ Bender and T.\ T.\ Wu,
                  Phys.\ Rev.\ D \underline{8} (1973) 3346;
                  T.\ Banks and C.\ M.\ Bender,
                  Phys.\ Rev.\ D \underline{8} (1973) 3366.

\bibitem{Jan}  W.\ Janke,
               Phys.\ Lett.\ A \underline{143} (1990) 107. 

\bibitem{KleTho}  H.\ Kleinert and S.\ Thoms,
                  Phys.\ Rev.\ D \underline{52} (1995) 5926.

\bibitem{FeKle}   R.\ P.\ Feynman and H.\ Kleinert,
                  Phys.\ Rev.\ A \underline{34} (1986) 5080.

\bibitem{GiToVa}  R.\ Giachetti and V.\ Tognetti,
                  Phys.\ Rev.\ Lett.\ \underline{55} (1985) 912;
                  Int.\ J.\ Magn.\ Matter.\ 
                  \underline{54}-\underline{57} (1986) 861;
                  R.\ Giachetti, V.\ Tognetti, and R.\ Vaia,
                  Phys.\ Rev.\ B \underline{33} (1986) 7647. 

\bibitem{book} H.\ Kleinert,
               {\em Path Integrals in Quantum Mechanics, Statistics 
               and Polymer Physics}, 
               2nd edition (World Scientific, Singapore, 1995)             

\bibitem{KaKle}  H.\ Kleinert,
                 Phys.\ Lett.\ B \underline{300} (1993) 261;
                 R.\ Karrlein and H.\ Kleinert,
                 Phys.\ Lett.\ A \underline{187} (1994) 133;
                 H.\ Kleinert and H.\ Meyer,
                 Phys.\ Lett.\ A \underline{184} (1994) 319,
                 see also: 
                 H.\ Kleinert,
                 Phys.\ Lett.\ A \underline{207} (1995) 133;
                 Phys.\ Lett.\ B \underline{360} (1995) 65.

\bibitem{JaKle}   W.\ Janke and H.\ Kleinert,
                  Phys.\ Lett.\ A \underline{199} (1995) 287.

\bibitem{SerConv} I.\ R.\ C.\ Buckley, A.\ Duncan, and H.\ F.\ Jones,
                 Phys.\ Rev.\ D \underline{47} (1993) 2554;
                 C.\ M.\ Bender, A.\ Duncan, and H.\ F.\ Jones,
                 Phys.\ Rev.\ D \underline{49} (1994) 4219;
                 A.\ Duncan and H.\ F.\ Jones,
                 Phys.\ Rev.\ D \underline{47} (1993) 2560;
                 R.\ Guida, K.\ Konishi, and H.\ Suzuki,
                 Ann.\ Phys.\ \underline{241} (1995) 152.

\bibitem{CoKleJa} H.\ Kleinert and W.\ Janke,
                  Phys.\ Lett.\ A \underline{206} (1995) 283;
                  R.\ Guida, K.\ Konishi, and H.\ Suzuki, Genova
                  preprint GEF-Th-4/1995 (hep-th/9505084)  

\bibitem{BeWu} C.\ M.\ Bender and T.\ T.\ Wu,
               Phys.\ Rev.\ \underline{184} (1969) 1231;
                Phys.\ Rev.\ D \underline{7} (1973) 1620.

\bibitem{Harm2} {\em Higher Transcendental Functions}, Bateman manuscript
                project, 
                ed.\ A.\ Erd\'{e}lyi (McGraw-Hill,
                New York, 1953) Vol.\ I. 

\bibitem{Abram} See, for example, 
                {\em Handbook of Mathematical Functions},
                edited by M.\ Abramowitz and I.\ A.\ Stegun
                (Dover, New York, 1965).

\bibitem{Hobson} E.\ W.\ Hobson, {\em The Theory of  
                 Spherical and Ellipsoidal Harmonics},
                 (Chelsea, New York, 1955).

\bibitem{Lang}  J.\ S.\ Langer,
                Ann.\ Phys.\ \underline{41} (1967) 108.

\bibitem{Lipat} L.\ N.\ Lipatov,
                JETP Lett.\ \underline{25} (1977) 104;
                L.\ N.\ Lipatov,
                Sov.\ Phys.\ JETP \underline{45} (1977) 216. 

\bibitem{Symanz} B.\ Simon and A.\ Dicke,
                 Ann.\ Phys.\ \underline{58} (1970) 76.  
\end{thebibliography}
\end{document}